\documentclass[fleqn,usenatbib]{mnras}

\usepackage{newtxtext,newtxmath}
\usepackage[T1]{fontenc}

\DeclareRobustCommand{\VAN}[3]{#2}
\let\VANthebibliography\thebibliography
\def\thebibliography{\DeclareRobustCommand{\VAN}[3]{##3}\VANthebibliography}

\usepackage{graphicx}	% Including figure files
\usepackage{amsmath}	% Advanced maths commands
\usepackage{comment}
\usepackage{subcaption}
\usepackage{ulem}
\usepackage{xcolor}

\title[High redshift H$_2$-DLAs]{Probing the physical environment of the most high-redshift H$_2$-DLAs through numerical models}

\author[ Shaji, Rawlins \& Kurel]{
Aashiya Anitha Shaji,$^{1,2}$\thanks{E-mail: aashiyashaji@gmail.com}
Katherine Rawlins,$^{2}$\thanks{E-mail: katherine.rawlins@gmail.com}
Pranshu Kurel$^{2}$
\\
$^{1}$LUX, Observatoire de Paris, Université PSL, Sorbonne Université, CNRS, 75014 Paris, France\\  
$^{2}$Department of Physics, St. Xavier's College (Autonomous), 5, Mahapalika Marg, Mumbai, 400001, Maharashtra, India}

\date{Accepted 2025 March 11. Received 2025 March 10; in original form 2023 July 31.}

\pubyear{2024}

\begin{document}
\label{firstpage}
\pagerange{\pageref{firstpage}--\pageref{lastpage}}
\maketitle

% Abstract of the paper
\begin{abstract}
%This is a simple template for authors to write new MNRAS papers.
%The abstract should briefly describe the aims, methods, and main results of the paper.
%It should be a single paragraph not more than 250 words (200 words for Letters).
%No references should appear in the abstract.
Damped Lyman-$\alpha$ absorbers (DLAs) with molecular hydrogen have been probed in detail through both spectroscopic observations and numerical modelling. However, such H$_2$ absorbers are quite sparse at very high redshifts. We identify six of the most distant known H$_2$-DLAs (redshift between 3 and 4.5), with medium/high-resolution spectroscopic observations reported in the literature, and perform detailed numerical modelling followed by Bayesian analysis to constrain their physical properties mainly using the H$_2$ rotational level population and \ion{C}{i} fine structure levels. Our modelling approach involves setting up a constant-pressure multiphase cloud irradiated from both sides, in comparison to most models which employ constant density. This enables us to use all observed atomic and molecular species as constraints to build a more realistic model of the DLA. Our results indicate high interstellar radiation field strength $\sim$ 10$^2$ to 10$^3$ G$_0$ for some sightlines, which is suggestive of \textit{in situ} star formation. The cosmic ray ionization rate for all DLAs is constrained between 10$^{-17}$ and 10$^{-14}$ s$^{-1}$, consistent with recent estimates for high-redshift sightlines. Total hydrogen density and temperature lie in the ranges 50 to 4 $\times$ 10$^4$ cm$^{-3}$ and 35--200 K in the innermost part of the absorbers. The corresponding gas pressure in our DLA models lies between 10$^{3.5}$ and 10$^{6.4}$ cm$^{-3}$ K, with three sightlines having a higher pressure than the range typical of high-redshift H$_2$-DLAs.
%Two of these DLAs are subject to enhanced X-ray radiation. For the other two absorbers, magnetic fields of strength 2--9 $\mu$G are needed to successfully reproduce the observed species column densities. These magnetic fields, akin to the mean Galactic interstellar field, are comparable to previous estimates for circumgalactic environments based on radio observations. Alternately, our results for these last two DLAs indicate that the observed H$_2$ excitation could be produced by additional heating in the absorber environment, pointing to the need for a modelling approach that includes dynamical effects.

\end{abstract}

% Select between one and six entries from the list of approved keywords.
% Don't make up new ones.

%%%%%%%%%%%%%%%%%%%%%%%%%%%%%%%%%%%%%%%%%%%%%%%%%%

%%%%%%%%%%%%%%%%% BODY OF PAPER %%%%%%%%%%%%%%%%%%

\section{Introduction}
Along the sightlines towards distant luminous sources like quasars, lie clouds of gas and dust that could arise from the intergalactic medium or the circumgalactic medium (CGM) of foreground galaxies and/or the host galaxy of the quasar itself \citep{Wolfe2005, Meiksin2009, Tumlinson2017}. The physical conditions within such quasar absorbers may be derived through analysis of their absorption lines in the spectrum of the background quasar. Depending upon the amount of neutral hydrogen, quasar absorbers are classified into damped Lyman-$\alpha$ systems (DLAs) with $N$(\ion{H}{i}) $>$ 10$^{20.3}$ cm$^{-2}$, sub-DLAs with 10$^{19}$ $<$ $N$(\ion{H}{i}) $<$ 2 $\times$ 10$^{20.3}$ cm$^{-2}$, Lyman limit systems (LLSs) with 10$^{17}$ $<$ $N$(\ion{H}{i}) $<$ 2 $\times$ 10$^{19}$ cm$^{-2}$, and the Lyman-$\alpha$ forest with $N$(\ion{H}{i}) $<$ 10$^{17}$ cm$^{-2}$ \citep{Wolfe2005, Meiksin2009}. 

%are routinely probed through the absorption lines they  leave behind

The neutral hydrogen column densities associated with DLAs are similar to column densities known within the Milky Way disc, thereby leading to the common understanding of DLAs as progenitors of modern-day spiral galaxies \citep{Storrie-Lombardie2000}. DLAs have been detected in the local Universe \citep{Oliveira2014, Muzahid2015, Kanekar2018, Dupuis2021}, as well as at high-redshift \citep{Prochaska2001, Noterdaeme2008, Dutta2014, Balashev2019, 2020Balashev}. Across the redshift interval, $\textit{z}=[0,5]$, DLAs are the most important reservoirs of neutral gas in the Universe \citep{Wolfe2005}. Constraining the physical conditions extant in DLAs over a broad redshift range will thus, help us understand the properties of gas associated with the formation and evolution of galaxies.

It was deduced that DLAs have a mean metallicity of 1/15 Z$_\odot$ at z $\sim$ 2--3, and dust-to-gas ratio 1/30 times that in the Galaxy \citep{Pettini1997}. However, there are also several instances of metal enrichment in high-redshift DLAs being comparable to that seen in the local Universe \citep{Noterdaeme2008b, Fynbo2011}. Like the phases of the Galactic interstellar medium, gas in DLAs can exist in the warm neutral \citep{Carswell2012} or cold neutral phases \citep{Wolfe2003b,Howk2005,Jorgenson2009}, with temperatures $\sim$ 50-16000 K \citep{2021Noterdaeme}, and hydrogen density $\sim$ 0.1-300 cm$^{-3}$ \citep{Draine2011, Klimenko2020}. Analysing high-resolution spectra of a sample of eighty DLAs, \citet{Neeleman2015} concluded that at least 5 per cent of DLAs exist in a cold phase with a temperature lower than 500 K and density $\sim$ 100 cm$^{-3}$. The typical pressure within these systems was inferred to be $\sim$ 2500 cm$^{-3}$ K.  

Some DLAs are cold enough to harbour molecules. While earlier studies revealed a 10--15 per cent detection rate of H$_2$ in high-redshift DLAs \citep{Noterdaeme2008}, more recent results indicate that the fraction could be as low as 4--6 per cent for N(H$_2$) > 10$^{17.5}$ cm$^{-2}$ \citep{Jorgenson2014, Balashev2018}. Alongside H$_2$, there are also a few instances of HD and CO detections \citep[eg.][]{Noterdaeme2008, Srianand2008, Noterdaeme2017, Kosenko2021}. H$_2$ column densities are often observationally constrained for several rotational levels of the ground vibrational state \citep{Albornoz2014, Noterdaeme2017, Rawlins2018}, based on the observed Lyman and Werner band transitions which involve absorption of photons with energy 11.2 to 13.6 eV. Besides UV photons, cosmic rays play an important role in determining the H$_2$ level population by pumping H$_2$ to the higher rovibrational levels \citep{Shaw2005, Shaw2006, Shaw2008}.

A significant fraction of H$_2$-DLA observational studies span the redshift range 2--3. This is mainly due to the atmospheric cut-off for ground-based UV observations which implies that H$_2$ can only be observed for DLAs beyond z$_{abs}$ $\sim$ 1.8. Further, the redshift interval of 2--3 coincides with the peaks of the quasar epoch and of the cosmic star formation history, allowing us to observe sightlines towards many bright background sources and thus, to probe an interesting epoch of galaxy evolution \citep{Richards2006, Ross2013, Madau2014}.

Neutral carbon has an ionization potential of 11.2 eV, comparable to the energies associated with the Lyman-Werner band transitions of H$_2$. Both species are mostly concomitant and associated with regions having the same temperature and density conditions in the cold neutral medium. Thus, absorption features of H$_2$ in various rotational levels are usually observed simultaneously with the fine structure lines of \ion{C}{i} \citep{Jorgenson2010, Rawlins2018}, and can be used to probe the properties of diffuse molecular gas. In the Galactic interstellar medium, thermal pressure was estimated based on observations of the \ion{C}{i} fine structure levels, and its median value was found to be $\sim$ 3800 cm$^{-3}$ K \citep{Jenkins2011}. This agrees well with the estimate of $\sim$ 2500 cm$^{-3}$ K for DLAs \citep{Neeleman2015}. Through a study of thirty-three high-redshift DLAs, \citet{Srianand2005} determined that the thermal pressure in those harbouring H$_2$ was in the range 824--30,000 cm$^{-3}$ K.

Column densities of multiple H$_2$ rotational levels or the fine structure levels of \ion{C}{i} (denoted as C {\sc i}, C {\sc i*}, and C {\sc i**}) are typically used to infer the radiation field, temperature and density properties of the gas under simple approximations \citep[][and references therein]{Noterdaeme2007, Albornoz2014, Rawlins2018}. But numerical models can yield better insight into the physical conditions in the absorbers with a detailed treatment of radiative transfer and chemistry. Thus, while molecular DLAs are relatively rare, they provide us with adequate observables to reconstruct the physical environment of such clouds through modelling studies. 

In recent years, several studies have adopted the approach of detailed numerical modelling to probe the physical environments of individual H$_2$-DLAs \citep{Shaw2016, Noterdaeme2017, Rawlins2018, Klimenko2020, Shaw2022}. While some of the modelled DLAs are at low-redshift \citep{Srianand2014}, there has been a great focus on redshifts $\left[ 2,3\right]$, keeping in line with the rich observations in this redshift interval. With current observational capabilities, there is limited insight into H$_2$-DLAs at greater distances. As a result, numerical models for such DLAs are rare for z$_{abs}$ $\gtrsim$ 3 \citep{Klimenko2020}. However, modelling absorbers at different redshifts is likely to give us more insight into their properties, and shed light on any evolutionary behaviour.

Based on medium/high-resolution spectroscopic observational studies in literature, we have identified six of the furthest H$_2$-bearing DLAs. In this paper, we investigate the physical conditions of these six absorbers. In Section \ref{sec:models}, we describe how the numerical models are set up using the software {\sc cloudy}, and present the best-fitting models for each of the six DLAs. The significance of our results is discussed in Section \ref{sec:disc}, followed by a summary in Section \ref{sec:conc}.

\section{Numerical models}
\label{sec:models} 
With the aim of studying H$_2$-DLAs located further than most systems modelled so far, we identified DLAs with z $\gtrsim$ 3 in literature with either H$_2$ and/or \ion{C}{i} reported. However, most of these sightlines either have low H$_2$ (N(H$_2$) (log cm$^{-2}$) < 18), or column density information available for only the lowest two H$_2$ rotational levels, making them unsuitable for rigorous modelling \citep{Ledoux2003, Balashev2014, Noterdaeme2018, Telikova2022}. Thus, we have selected the following six H$_2$-DLAs for further study in this paper: (i) DLA at $z$ = 3.033 towards Q J1236+0010, (ii) DLA $z$ = 3.091 towards Q J2100-0600, (iii) DLA at $z$ = 3.093 towards the quasar J1311+2225, (iv) DLA at $z$ = 3.255 towards the quasar J2205+1021, (v) DLA at $z$ = 3.287 towards the quasar SDSS J081634+144612, and (vi) DLA at $z$ = 4.224 towards PSS J1443+2724. Further details about the modelling methodology and observations reported in literature for each sightline are presented in the corresponding sub-sections below, along with the physical properties constrained by our numerical models.

\subsection{{\sc cloudy} and the modelling approach}
\label{ssec:method} 

We use the spectral synthesis code {\sc cloudy} version 23.01 \citep{Ferland2017, 2023Chatzikos, 2023Gunasekera} to model the six DLAs by solving for the temperature, radiation field, and abundances of atomic and molecular species, as a function of depth into the cloud. When ultraviolet radiation is incident on a plane-parallel slab of gas, a photodissociation region (PDR), that is, a region dominated by photons of energy 6 eV < $h\nu$ < 13.6 eV, is set up near the illuminated face of the cloud \citep{Tielens1985}. In this study, we use the structure of a typical PDR to model the high-redshift H$_2$-bearing DLAs. {\sc cloudy} performs self-consistent calculations that determine the chemical, thermal and ionization balance at each point within the cloud. The model absorber is divided into various zones of a finite thickness that is small enough for the physical conditions to be assumed as constant throughout the particular zone. {\sc cloudy} accounts for various radiative and collisional processes and calculates column densities and line intensities corresponding to several atomic and molecular species. The 30 lightest elements and several molecules are included in the {\sc cloudy} chemistry network \citep{Lykins2015, Ferland2017}. Various H$_2$-related processes and reactions are implemented in detail within {\sc cloudy} \citep{Shaw2005}. Extensive treatment of dust grain physics is also incorporated in {\sc cloudy} \citep{Hoof2004}. This is particularly relevant to H$_2$, as the molecule is mainly formed on grain surfaces in metal-enriched environments.

We adopt a two-sided radiation approach in our models, with the metagalactic background radiation due to galaxies and quasars \citep{2019}, henceforth referred to as KS19, the redshift-appropriate cosmic microwave background (CMB), and an interstellar radiation field due to \textit{in situ} star formation, impinging both faces of the plane-parallel geometry of the absorber. In order to execute this, one-half of the cloud is modelled corresponding to half the observed value of N(H$_2$). The predicted column densities are doubled in order to reproduce the total column densities for the entire absorber. The total pressure is kept constant throughout the extent of the modelled cloud, and its value is determined self-consistently by {\sc cloudy} during the calculations. In addition to the thermal motion of atoms, local non-thermal effects are accounted for through a micro-turbulent velocity. Observational constraints are usually available for this, through the Doppler parameter determined during analysis of observed absorption line profiles.

One of the key parameters in our models is the total density of hydrogen nuclei ($n_H$) at the illuminated face of the cloud. This represents the sum of hydrogen nuclei in all forms, that is mainly, atomic ($n_{HI}$), ionized ($n_{HII}$), and molecular hydrogen ($n_{H_2}$) and is expressed as:
\begin{equation}
    n_H = n_{HI} + n_{HII} + 2 n_{H_2} + \Sigma n_{H_{\text{other}}}
\end{equation}
This parameter specified at the illuminated surface of the cloud serves as one of the primary variables used to describe the physical conditions within the absorber. In our models, $n_H$ varies as a function of depth into the cloud, reflecting changes in the physical and chemical conditions as calculated self-consistently by {\sc cloudy}, which also accounts for thermal and pressure balance.

The chemical composition of the DLA clouds is specified in terms of solar abundances from \citet{Grevesse2010}. The corresponding observational estimate of metallicity is applied to the models along with the expected depletion behaviour of different elements on to dust grain surface according to \citet{2009Jenkins}. In {\sc cloudy}, the overall depletion is described by a single parameter, the depletion strength (denoted as $F_*$), which provides an empirical relationship between the individual depletion scale factors of different elements. We conform to the default value of \( F_* = 0.5 \), which represents the median depletion level in the \citet{2009Jenkins} sample, as it corresponds to sight lines typical of diffuse \ion{H}{II} regions with average densities. This choice is consistent with the approach used by \citet{2021Hensley} and \citet{2022Gunasekera} ensuring that the depletion pattern is representative of the diffuse ISM, where metal depletion into dust grains is moderate. Dust grains with properties similar to that of the Galactic interstellar medium (ISM) have been employed. Both graphites and silicates follow the grain size distribution of \citet{Mathis1977}:
\begin{equation}
    \frac{dn}{da} \propto a^{-3.5}.
\end{equation}
Here, $n$ is the number of grains per unit volume and a is the dust grain radius lying within the range $0.005-0.250 \mu$m.

The dust-to-gas ratio is typically calculated using observed metal column densities. If this value is not already reported in the literature, we obtain an estimate using the standard relation of \citet{Wolfe2003b}:
\begin{equation}
    \kappa = 10^{[X/H]} (1 - 10^{[Y/X]}).
\end{equation}
Here, X is a volatile species and Y is a refractory species. The value obtained thus, provides an estimate of the dust abundance to be considered in the DLA model. We calculated the dust-to-gas ratio using this method for the DLAs at z = 3.033, 3.091, 3.255 and 3.287. 
%In both cases, the volatile species was zinc while the refractory species was iron. \textcolor{green}{Pranshu details}

%The cosmic ray ionization rate is understood to have a mean value of 2 $\times$ 10$^{-16}$ s$^{-1}$, as determined by \citet{Indriolo2007} using observations of H$^{+}_3$ along several Galactic sightlines. The cosmic ray ionization rate is not yet very well understood for high-redshift galaxy environments. We incorporate this Galactic mean in our models and then enhance or suppress the rate, as found necessary to explain the observed $N$(\ion{H}{i}) and H$_2$ rotational level population.

We incorporate the effects of cosmic ray ionization in our models by specifying the ionization rate. This is found necessary to explain the observed $N$(\ion{H}{i}) and H$_2$ rotational level population in diffuse gas in the Galaxy \citep{,Shaw2006, Shaw2008, 2009Padovani}. A mean value of 2 $\times$ 10$^{-16}$ s$^{-1}$ was determined by \citet{Indriolo2007} using observations of H$^{+}_3$ along several Galactic sightlines. However, the cosmic ray ionization rate is not yet very well understood for high-redshift galaxy environments.

Overall, our models have the following parameters -- total density of hydrogen nuclei at the illuminated face ($n_H$), shape and intensity of the interstellar radiation field (ISRF: $I_{UV}$), and cosmic ray ionization rate (CRIR: $\zeta$). We determine ISRF in the units of Habing field \citep[][$G_0 = 1.6 \times 10^{-3} \text{erg cm}^{-2}\text{s}^{-1}$ between 6 and 13.6 eV]{Habing1968, Tielens1985}. The KS19 metagalactic background is included separately, with the flux contribution depending on the absorber redshift. We adopt a methodology that jointly constrains physical parameters and their uncertainties \citep{Neeleman2015, Klimenko2020, Kosenko2021, 2022Acharya}.

For each of the six DLAs, we run a grid of {\sc cloudy} simulations over log $n_H$ = -3 to +3 ($\Delta$log $n_H=0.3$), log $I_{UV}$ = -1 to 3 ($\Delta$log $I_{UV}=0.5$), and log $\zeta$ = -17 to -14 ($\Delta$log $\zeta$=0.5). We choose a finer sampling of $n_H$ to better resolve the effects of density variations in our isobaric models. Model-predicted column densities of various observed species are stored for each parameter combination $\theta$ (log $I_{UV}$, log $n_H$, log $\zeta$). We employ Bayesian formalism with the Markov Chain Monte Carlo (MCMC) technique, using the {\sc emcee} library \citep{2013Foreman-Mackey}, to explore the posterior probability distribution function (PDF) for each parameter ($\theta$). We assume a Gaussian likelihood function where the uncertainty on the observed column densities defines the standard deviation in the likelihood equation. Flat priors are used within the parameter ranges covered by our grids. The posterior probability is then computed as the product of the likelihood and prior, in accordance with Bayes' theorem. Using MCMC, we sample the posterior PDF by linearly interpolating between the {\sc cloudy} model grid points. From the resulting posterior distributions, we plot the corner plots (in Figs. \ref{fig:corner_3.287}, \ref{fig:corner_3.093}, \ref{fig:corner_3.091}, \ref{fig:corner_3.255}, \ref{fig:corner_4.224}, and \ref{fig:corner_3.033}) determine the most likely value for each parameter as the median of the distribution. The $16^\text{th}$ and $84^\text{th}$ percentiles are used to represent the 1-sigma uncertainties.

\begin{table*}
    \centering
    \caption{Details of the modelled DLAs. The parameters log I$_{UV}$, log n$_H$ and log $\zeta$ are constrained through {\sc cloudy} models and Bayesian analysis. Other model parameters are based on observational estimates (references are provided in-text).}
    \begin{tabular}{ccccccccc}
        \hline
        Quasar name & DLA redshift &  log N(H$_2$) (in cm$^{-2}$) & log I$_{UV}$ (in G$_0$) & log n$_H$ (in cm$^{-3}$) & log $\zeta$ (in s$^{-1}$) & Z & $\kappa$ & b (km s$^{-1}$)\\
        \hline
        Q J1236+0010 & 3.033 & 19.76 $\pm$ 0.01 & $2.54 \pm 0.02$ & $2.97^{+0.02}_{-0.04}$ & $-14.01^{+0.01}_{-0.02}$ & $-0.58^{+0.04}_{-0.03}$ & 0.24 & 2.3 $\pm$ 0.03\\
        Q J2100-0600 & 3.091 & 18.76 $\pm$ 0.01 & $1.55^{+0.02}_{-0.03}$ & $1.25^{+0.27}_{-0.11}$& $-14.50^{+0.42}_{-0.03}$ & -0.73 $\pm$ 0.15 & 0.17 & 2.3\\
        J1311+2225 & 3.093 & 19.59 $\pm$ 0.01 & $2.78^{+0.12}_{-0.26}$ & $0.62^{+0.05}_{-0.02}$ & $-14.54 \pm 0.03$ & $-0.34^{+0.13}_{-0.14}$ & 0.46 & 5.4\\   
        2205+1021 & 3.255 & 18.16 $\pm$ 0.03 & $2.67^{+0.16}_{-0.27}$ & $1.21^{+0.30}_{-0.44}$ & $-14.99^{+0.34}_{-1.34}$ & -0.93 $\pm$ 0.05 & 0.1 & 1\\
        081634+144612 & 3.287 & 18.66 $\pm$ 0.87 & 2.01 $\pm$ 0.03 & 1.03 $\pm$ 0.04 & -16.99 $\pm$ 0.01 & -1.10 ± 0.10 & 0.05 & 4.7 \\
        PSS J1443+2724 & 4.224 & 18.28 $\pm$ 0.09 & $1.03^{+0.10}_{-0.05} $ & $0.00^{+0.02}_{-0.35}$ & $-15.91^{+0.02}_{-0.12}$ & -0.63 $\pm$ 0.10 & 0.16 & 1.2 $\pm$ 0.6\\
        \hline
    \end{tabular}
    \label{tab:params}
\end{table*}

\begin{comment}
\sout{These parameters are suitably varied over a physically acceptable range of values until the column densities predicted by the numerical models match the observed ones to within a factor of two, that is, 0.3 dex, or to within the observational uncertainty if that is larger. While {\sc cloudy} offers an in-built PHYMIR algorithm for model optimization \citep{vanHoof1997, Ferland2013}, this works towards $\chi^2$-optimization for the overall model. Often, this results in some observables being very well-constrained while some others are in poor agreement, as also discussed in \citet{Rawlins2018}. We thus obtain our final best-fitting DLA models through a manual approach of fine-tuning the physical parameters, after initially surveying the parameter space through optimization runs or by setting up grids of models. The minimum interval by which we vary each physical parameter is expressed as the uncertainty in the constrained value.}
\end{comment}

The physical parameters constrained for all six DLAs through the Bayesian analysis are given in Table \ref{tab:params}, along with some of the other model parameters taken from observational estimates. We use the constrained parameters to run best-fitting {\sc cloudy} models for each of the six systems, allowing us to determine their physical conditions as a function of depth. In the following sub-sections, we discuss more details of the insight from our six modelled DLAs. Since {\sc cloudy} does not inherently provide uncertainty estimates on predicted physical properties, we generate additional {\sc cloudy} models for each DLA corresponding to the maximum and minimum values of all physical parameters (all possible permutations), as obtained from the Bayesian analysis. Thus, considering the three input parameters (I$_{UV}$, n$_H$, $\zeta$) leads to eight additional models per DLA. As there are non-linear dependencies, we explore all permutations in order to capture the full range of resultant variation in various physical properties of each system. In the sub-sections below, we mainly report the results arising from the model corresponding to the central parameter values, but use the models with the extreme values of the parameters to provide uncertainty/range estimates.

\begin{comment}
The physical parameters that were supplied to the best-fitting models of all six DLA systems are tabulated in Tables \ref{tab:DLA1a}, \ref{tab:DLA4a}, \ref{tab:DLA5a}, \ref{tab:DLA3a}, and \ref{tab:DLA2a} \textcolor{cyan}{Instead of Table 1 and all these tables, one table with the 3 quantities constrained from emcee, metallicity, kappa, b-parameter should be sufficient. In the corresponding sections, details of observed quantities and contribution of KS19 in terms of G0 could be pointed out (you'll find this in the .out file).}. \sout{The observed and model-predicted column densities are tabulated in Tables \ref{tab:DLA1b}, \ref{tab:DLA4b}, \ref{tab:DLA5b}, \ref{tab:DLA3b}, and \ref{tab:DLA2b}.}
\end{comment}

\subsection{DLA towards J081634+144612 ($z_{abs}=3.287$)}
\label{ssec:dla1}

The DLA at $z=3.287$, lying in the foreground of quasar SDSS J081634+144612 ($z_{em} = 3.84$), was observed by \citet{Guimaraes2012} using the Ultraviolet and Visual Echelle Spectrograph (UVES) on Kueyen, the second 8.2-m Unit Telescope (UT2) of the ESO Very Large Telescope (VLT) at Chile at a spectral resolution of 50,000. Based on H$_2$ excitation, the authors have calculated the kinetic temperature to be $\sim$ 75 K, while the total hydrogen density in the molecular region was estimated to lie in the range 50--80 cm$^{-3}$ using the column density ratios of the \ion{C}{i} levels. Based on their analysis of the observed spectrum, the authors have concluded that around 90 per cent of the total H$_2$ in the DLA is found to coincide with the \ion{C}{i} component, which is consistent with expectation \citep{Guimaraes2012}. Absorption features of five singly ionized species are also detected -- \ion{Zn}{ii}, \ion{Fe}{ii}, \ion{Cr}{ii}, \ion{Ni}{ii} and \ion{Si}{ii} -- spread across 150 km s$^{-1}$. Though metal species were resolved into eight different velocity components, a few of them were clumped together, with redshifts of 3.28661, 3.28735 and 3.28814. H$_2$-absorption is detected in two of these three clumps. Here, we try to model these components as a single cloud. Thus, the sum of column densities for each absorbing species over these individual components is considered as the total observed value.

\begin{figure}
    \centering
    \includegraphics[width=\linewidth]{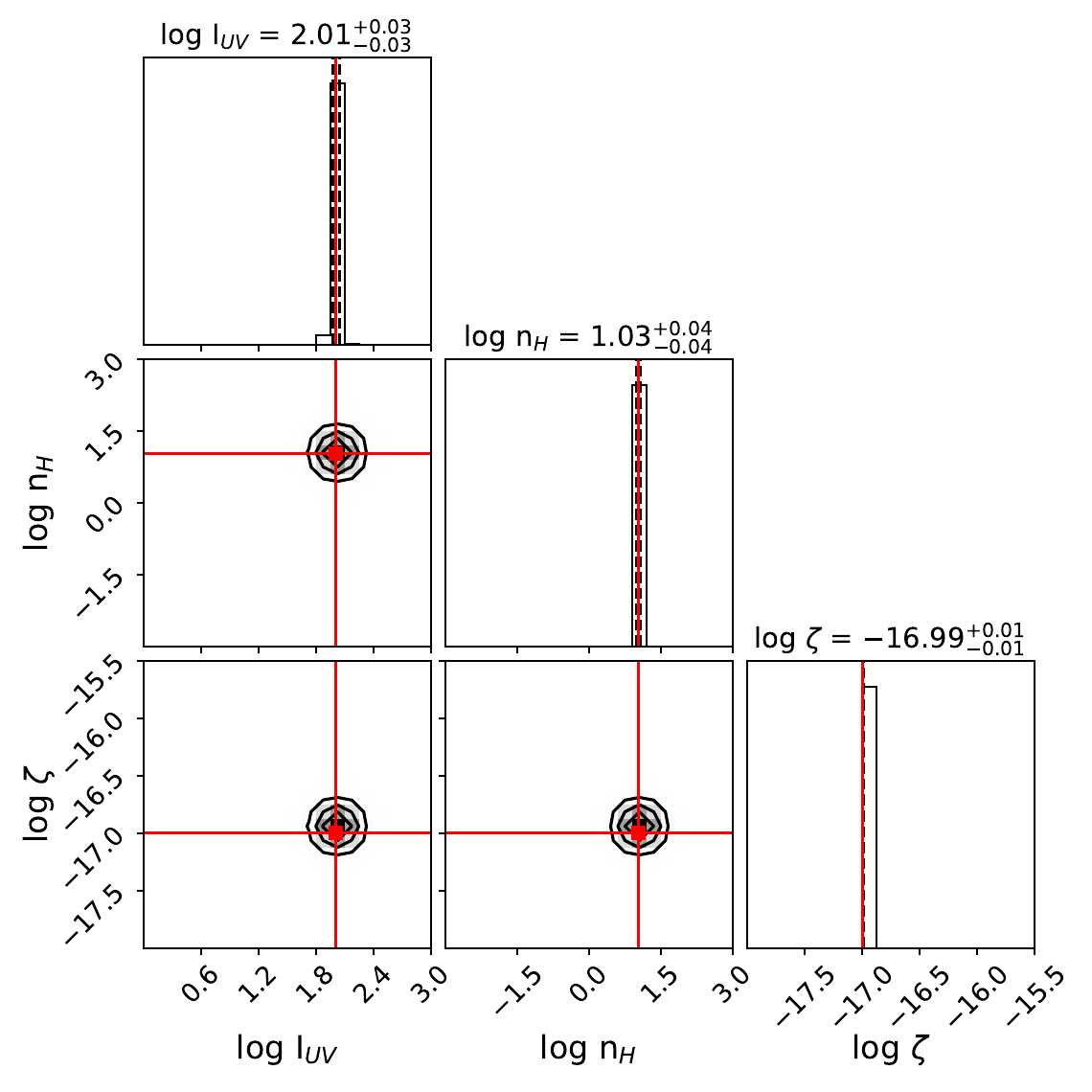}
    \caption{Joint and marginalised probability distributions of log I$_{UV}$ (in $G_0$), log n$_H$ (in cm$^{-3}$) and log $\zeta$ (in s$^{-1}$) for the DLA towards J081634+144612 ($z_{abs}=3.287$). Here, I$_{UV}$ refers only to ISRF, excluding the metagalactic background (incorporated separately). The red squares and lines indicate the maximum and median posterior values, respectively, for each parameter. The dashed lines represent the 16$^\text{th}$ and 84$^\text{th}$ percentiles, indicating the 1$\sigma$ uncertainty ranges.}
    \label{fig:corner_3.287}
\end{figure}

\begin{comment}
\begin{table}
    \caption{Input parameter values of the best-fitting model for the DLA at $z_{abs} = 3.287$ }
        \centering
        \begin{tabular}{lcc} % four columns, alignment for each
	    \hline
        Physical parameter & Model value  & Observed value\\
        \hline
        Radiation field & KS19+CMB+ISRF & -\\
        Total UV Intensity & 1.35 $\pm$ 0.10 G$_0$ & -\\
        UV Intensity of KS19 & 0.31 G$_0$ &-\\ 
        UV Intensity of ISRF& 1.04 $\pm$ 0.10 G$_0$ &-\\ 
        Density at illuminated face, n$_{\rm H}$ & 0.4 $\pm$ 0.2 log cm$^{-3}$ & 50-80 cm$^{-3}$\\ 
        & & (in molecular region)\\
        Metallicity, {[Zn/H]}   & -0.85 $\pm$ 0.05 & -1.10 $\pm$ 0.10\\
        {[C/H]} & -1.95 $\pm$ 0.05  & - \\
        {[Ni/H]} & -1.38 $\pm$ 0.15 & -1.68 $\pm$ 0.10\\
        {[Fe/H]} & -1.31 $\pm$ 0.12 &  -1.58 $\pm$ 0.10\\
        {[Cr/H]} & -1.25 $\pm$ 0.10 &  -1.58 $\pm$ 0.10\\
        Dust-to-gas ratio, $\kappa$ & 0.04 $\pm$ 0.01 & 0.05\\ 
        Micro-turbulence & 4.5 $\pm$ 0.2 km s$^{-1}$ & 4.7 km s$^{-1}$\\
        Cosmic ray ionization rate & -15.25 $\pm$ 0.05 log s$^{-1}$ & -\\
        \hline
        \end{tabular}
        \label{tab:DLA1a}
\end{table}  
\end{comment}

\begin{figure*}
	\includegraphics[width=\textwidth, keepaspectratio]{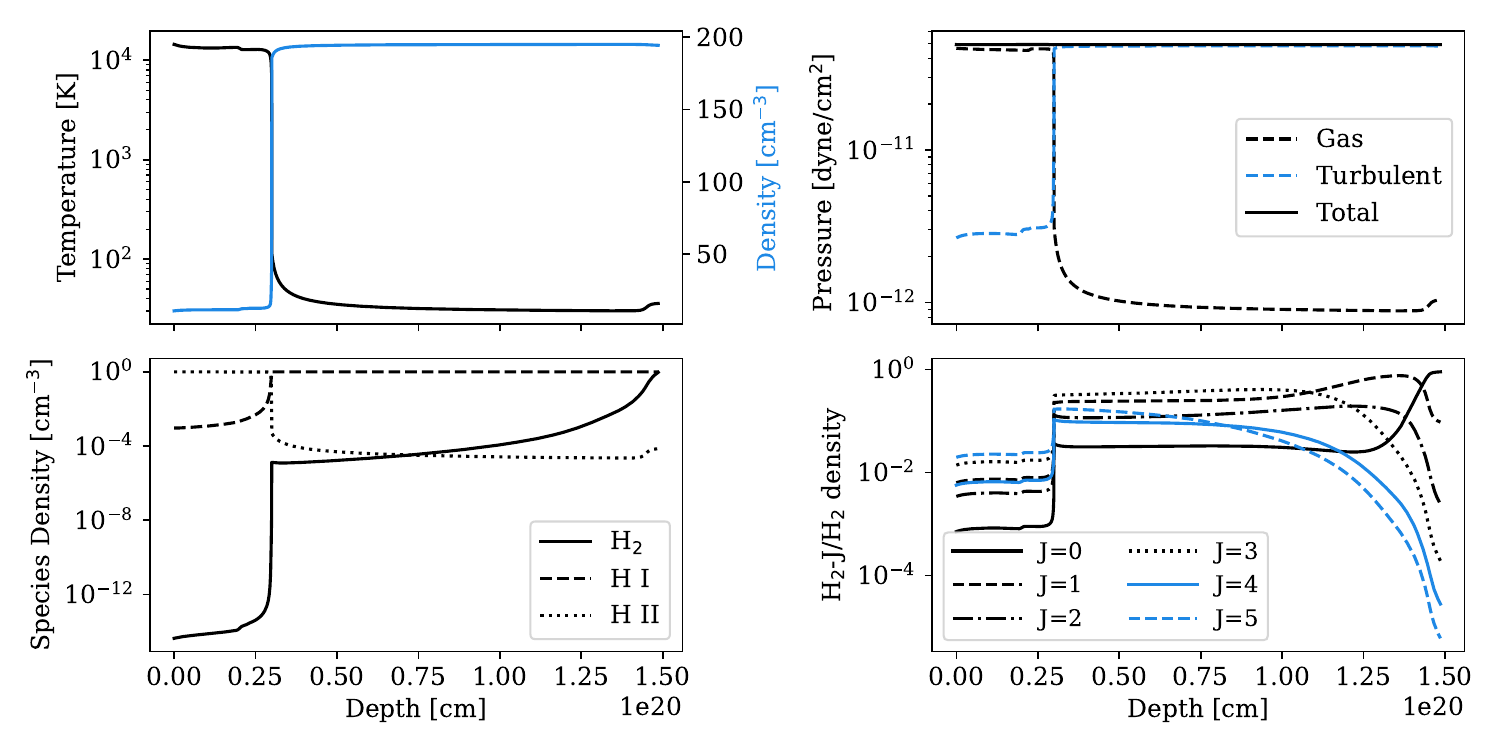}
    \caption{Physical conditions within one half of the DLA towards SDSS J081634+144612 at $z_{abs} = 3.287$ plotted based on input values of $\log I_{UV} = 2.01$, $\log n_H = 1.03$ and $\log \zeta = -16.99$. The other half of the cloud bears a symmetric profile. The different panels plot the following as a function of depth into the cloud from the illuminated face: (i) Density and temperature, (ii) Pressure and its constituents, (iii) Species density, (iv) Density of individual H$_2$ rotational levels, normalised to the total H$_2$ density.}
    \label{fig:DLA1}
\end{figure*}

From the Bayesian analysis (as seen in Fig. \ref{fig:corner_3.287}), the total radiation field, hydrogen density and the cosmic ray ionisation rate at the illuminated (incident) face were set at 2.01 log G$_0$, 1.03 log cm$^{-3}$ and -16.99 log s$^{-1}$ respectively. The metallicity, dust-to-gas ratio and microturbulent velocity were set according to the observed values from \citet{Guimaraes2012} which have been tabulated in Table \ref{tab:params}. 
%They were then varied suitably to reproduce the column densities as well as possible. The values that constrain our best-fitting models are consistent with the observational estimates and have been listed in Table \ref{tab:DLA1b}. %\sout{The elemental abundances for C, Ni, Fe and Cr are varied separately from the overall metallicity since the column densities of their ions are known. The abundances of these species are also likely to affect the H$_2$ rotational level population due to collisional processes. We discuss this further in Section \ref{ssec:metals} \citep[see also][]{Shaw2016, Rawlins2018}.}
%Almost all column density predictions in the best-fitting model lie within a factor of 2 of the observed values, or within the observational uncertainties where these are larger. Only for the H$_2$ $J = 2$ rotational level, the column density from the model is significantly higher than that observed. 

Fig. \ref{fig:DLA1} represents the variation of properties with depth within one-half of the cloud. The other half then follows a symmetrical profile. The properties shown in the top panel are the temperature of the gas and total hydrogen density (upper left panel), and the constituents making up the total pressure (upper right panel). The total (atomic, ionic and molecular) hydrogen density at the illuminated face of the cloud is 1.03 log cm$^{-3}$. As the cloud becomes cooler with increasing depth from the illuminated face, the total hydrogen density increases and reaches a value of 200 cm$^{-3}$ in the inner regions of the absorber. Considering the uncertainties in the input parameter values as constrained by {\sc emcee}, this density in the interior of the DLA may go up to 1400 cm$^{-3}$. Out of the two components contributing to the total pressure of the system, the gas pressure governs in the hotter regions and the pressure due to microturbulence dominates in the colder region, as was also reported by \citet{Rawlins2018}. The density conditions are consistent with those expected for the cold neutral medium, and the gas pressure in the innermost part of the cloud is 7500 cm$^{-3}$ K in our representative model. However, when considering the full range of parameter uncertainties, the gas pressure can lie between 8000 and 43,000 cm$^{-3}$ K. This remains towards the higher range of pressure values known for H$_2$-DLAs, i.e. 824--30,000 cm$^{-3}$ K \citep{Srianand2005}.

The bottom row of Fig. \ref{fig:DLA1} plots the trends in the density of atomic, ionic and molecular hydrogen (lower left panel), and the density of the H$_2$-rotational levels (lower right panel). In keeping with expectation, there is an evident shift from ionic to atomic to molecular contents in the gas, as we move deeper into the cloud. However, the cloud does not become fully molecular even in the coolest regions, continuing to be dominated by \ion{H}{i}. The population trend of the rotational levels in the ground vibrational state of molecular hydrogen is closely related to the formation of H$_2$ molecules in the cloud. The population in the J = 0 rotational level increases with depth and becomes the dominant form of H$_2$ beyond the \ion{H}{ii}-\ion{H}{i} interface. Due to shielding in the cloud interiors, fewer photons are available there to excite H$_2$ molecules onto higher energy states ($J=3,4,5$).

The heating caused by photoionization of \ion{H}{i} dominates in the atomic region, whereas grain photoelectric heating followed by cosmic-ray heating becomes more prominent while moving towards the interior, where there is a paucity of UV photons. There is a significant contribution of \ion{He}{i} photoionization and heating due to charge transfer across the ionization front. However, this gradually drops off with increasing depth into the absorber as the incident UV photons become attenuated. 

\begin{comment}
    Cooling in the neutral parts of the cloud occurs mainly through collisional line cooling due to hydrogen, with maximum contributions from [\ion{C}{ii}] 158 $\mu$m lines in the molecular region.
\end{comment} 

\subsection{DLA towards J1311+2225 at $z_{abs}=3.093$}

H$_2$ was reported for the first time along this sightline by \citet{Noterdaeme2018} using high-resolution spectra obtained using UVES at VLT. They also detect \ion{C}{i} absorption associated with the same DLA. The absorption lines were analysed again by \citet{Kosenko2021} during their study of HD molecules and cosmic ray ionization at high-redshift. We use their updated H$_2$ column density estimate of 19.59$\pm$0.01 as the observational constraint for our modelling. \citet{Kosenko2021} determined the metallicity of the absorber using \ion{Zn}{ii} absorption. It was found to be -0.34$^{+0.13}_{-0.14}$. For our models, we assume a dust abundance approximately similar to the metal abundance of the system. \citet{Noterdaeme2018} do not detect CO towards this sightline, and estimate that N(CO) $<$ 13.43 cm$^{-2}$. Consistent with this, our model predicts N(CO) = 10.3 cm$^{-2}$.

\begin{comment}
Along with Bayesian analysis using our grid of {\sc cloudy} models with a range of I$_{UV}$, n$_H$, and $\zeta$ values, 
    the optimum model was decided by also inspecting the size of the predicted cloud. Higher intensities of ISRF and CRIR produced larger absorbers, so the highest values were excluded from the priors for this DLA.
\end{comment} 

This system is modelled using an incident intensity of 2.78 log G$_0$, total hydrogen density of 0.62 log cm$^{-3}$ and cosmic ray ionisation rate of -14.54 s$^{-1}$ at the incident face. The ionic region of the cloud is characterized by a low hydrogen density and a high temperature ($ 10^4$ K), as shown in Fig. \ref{fig:3.093_physical}. The \ion{H}{ii}--\ion{H}{i} transition occurs at $3\times10^{20}$ cm,  which, in this particular DLA seems to coincide with the over-production of H$_2$. Here, the total hydrogen density increases to 40--60 cm$^{-3}$, and the temperature drops to 150--170 K simultaneously. The gas pressure in this interior region lies in the range $(7.4-9.7) \times 10^3$ cm$^{-3}$ K. The density of H$_2$ barely overtakes that of \ion{H}{i} which is the indication of a relatively warm, diffuse area as compared to the more molecular environments seen elsewhere.

The molecular hydrogen abundance slightly exceeds that of atomic hydrogen in the cloud's inner regions. As in previous cases, the $J=0,1$ levels remain heavily populated. Heating in the illuminated phase is dominated by \ion{He}{ii} photoionization, while cosmic rays take over deeper into the cloud. Cooling is primarily due to emissions from oxygen in the outer regions, while [\ion{C}{ii}]  158 $\mu$m lines contribute significantly in the denser, central areas.

\subsection{DLA towards Q J2100-0600 at ($z_{abs}=3.091$)}

Observations for the DLA towards towards Q J2100-0600 at z$_{abs}$ = 3.091 have been reported in \citet{Balashev2015} and \citet{Jorgenson2010}. The emission redshift associated with the background quasar is z$_{em}$ = 3.14. H$_2$ is detected in the lowest six rotational levels of the ground vibrational state. A metallicity value of -0.73$\pm$0.15 is estimated for this DLA based on Si absorption. 
The Bayesian analysis of this DLA indicates that the model-predicted column densities best match the observed ones when the incident $\log I_{UV} = 1.55 G_0$, $\log n_H = 1.25 \text{cm}^{-3}$ and $\log \zeta = -14.50 \text{ s}^{-1}$. The gas temperature and hydrogen density in the interior of this cloud are estimated to lie between 50–150 K and 600–2000 cm$^{-3}$, with the estimates corresponding to the central best-fitting parameter values being 50 K and 760 cm$^{-3}$. The estimated rotational excitation temperature for this line of sight is 159$^{+44}_{-29}$ K \citep{Klimenko2020}, which is in agreement with our values. The gas pressure in the innermost part of the absorber is approximately 33,400 cm$^{-3}$ K, falling within the larger estimated range of $3.3 \times 10^4$ to 1.4 $\times 10^5$ cm$^{-3}$ K given by the extreme values of the three input parameters. This sightline has previously been modelled by \citet{Klimenko2020} and we provide a comparison with their results in Section \ref{ssec:model-comp}.

Fig. \ref{fig:DLA_3.091} illustrates how various physical parameters evolve within the modeled absorber. The \ion{H}{ii}-\ion{H}{i} transition occurs at a shallow depth of $ 2\times 10^{17}$ cm from the illuminated face, while the \ion{H}{i}-H$_2$ transition takes place deeper, at $4 \times 10^{17}$ cm. The latter marks a sharp shift in temperature, density, and pressure constituents, signifying the transition from the warm neutral to the cold molecular phase. Deeper into the cloud, the H$_2$ number density eventually surpasses that of \ion{H}{i}. As expected, higher H$_2$ rotational levels are populated closer to the illuminated face, while lower levels become prominent only beyond the \ion{H}{i}-H$_2$ transition.

The heating near the illuminated surface is due to \ion{H}{i} photoionization but cosmic rays and \ion{He}{i} photoionization become important in the inner reaches of the cloud. Grain photoelectric heating is relatively weak throughout this absorber. The cooling is mostly through the collision channel of H atoms in the illuminated phase and [\ion{C}{ii}] the emission of 158 $\mu$ m dominates 80\% of the cooling in the molecular part of the cloud.

\subsection{DLA towards J2205+1021 ($z_{abs}=3.255$)}
This DLA lies towards the quasar J2205+1021 (z$_{em}$=3.41), and was observed using the multi-wavelength, medium-resolution X-Shooter at VLT, Chile \citep{Telikova2022}. Metal and \ion{H}{i} column densities for this DLA have been reported in literature \citep{Telikova2022}, while the H$_2$ and \ion{C}{i} column densities were obtained through private communication with the lead author of the study. The metallicity is -0.93, and the dust-to-gas ratio is 0.1 times the standard ISM value.

%In this DLA too, the observed H$_2$ rotational level population cannot be reproduced simultaneously with the column density of \ion{H}{i}. As for the previously discussed DLAs, the incident radiation field, cosmic ray ionization rate, total hydrogen density, metallicity, dust-to-gas ratio, and microturbulent velocity were varied suitably about the corresponding observationally deduced estimates. However, none of the models reproduced the large column densities in the higher H$_2$ rotational levels even when the cosmic ray ionization rate was significantly enhanced over the Galactic mean. Only a model employing a magnetic field of strength 2.5 $\mu$G near the irradiated face of the absorber proves to be effective. The predicted column densities of mostly all species either lie within a factor of 2 of the corresponding observed ones or fall within the observational uncertainties if they are larger (see Table \ref{tab:DLA3b}). Only the population of the H$_2$ J = 3 level in the model is lower than the observed values. The cosmic ray ionization rate considered in this model cloud is similar to the mean Galactic background value of 2 $\times$ 10$^{-16}$ s$^{-1}$ \citep{Indriolo2007}. The average value of the magnetic field in the neutral interstellar gas of our Galaxy is 6 $\mu$G \citep{Heiles2004} and is comparable to the value we constrain here for this high-z DLA. We discuss the significance of the magnetic field further in Section \ref{sec:disc}.  

Fig. \ref{fig:3.255_physical} depicts the variation of different properties as a function of depth into one-half of the cloud. The input parameters of such a model have been listed in Table \ref{tab:params}. The top panels illustrate the fall in temperature and increase in total hydrogen density with depth, along with an insight into the constituents of pressure. The temperature and gas density in the innermost region of the cloud are found to be 70 K and $3.5 \times 10^3$ cm$^{-3}$, respectively. Considering the uncertainties in our input physical parameters, these predicted values lie within a larger estimated range of 40–90 K and $(1.1-7.6) \times 10^3$ cm$^{-3}$. In the most shielded part of the absorber, the gas pressure is $2.5 \times 10^5$ cm$^{-3}$ K, falling within the broader range of $7.4 \times 10^4 - 5.2 \times 10^5$ cm$^{-3}$ K. The gas remains dense and cold in the inner parts of the cloud.

From the abundance plots, it can be seen that the transition from \ion{H}{i} to H$_2$ has not fully occurred. Heating occurs predominantly through photoionization of \ion{H}{i} and \ion{He}{i}. In the interior regions, grain photoelectric heating also acquires some importance,  though it remains secondary to \ion{H}{i} photoionization. Cooling mainly occurs through the channels of oxygen and carbon emission.
%Line emission through \textcolor{green}{[OIII], [O II]  idk wavelengths} and [\ion{C}{ii}]  158 $\mu$m act as the major cooling agents throughout the cloud. %H$_2-J=3,4$ can be better reproduced by employing a density gradient throughout the cloud.} \textcolor{green}{please check this statement}

\begin{comment}
\begin{table}
    %\caption{DLA at $z_{abs} = 3.255$}
    %\begin{subtable}{.5\linewidth}
      \centering
        \caption{Input parameter values of the best-fitting model for the DLA at $z_{abs} = 3.255$ }
        \begin{tabular}{lcc} % four columns, alignment for each
	\hline
    Physical parameter & Model value & Observed value \\
    \hline
    Radiation field & KS19+CMB+ISRF & --\\
    Total UV Intensity & 1.35 $\pm$ 0.5 G$_0$  & -\\
    UV Intensity of KS19 & 0.31 G$_0$ &-\\ 
    UV Intensity of ISRF& 1.04 $\pm$ 0.10 G$_0$  &-\\ 
    Density at illuminated face, n$_{\rm H}$  & -0.01 $\pm$ 0.20 log cm$^{-3}$ & --\\ 
    Metallicity, {[Zn/H]}   & -0.92 $\pm$ 0.05 & -0.93 $\pm$ 0.05\\
    {[C/H]} & -1.35 $\pm$ 0.14 & --\\
    {[Ni/H]} & -1.71 $\pm$ 0.10 & -1.69 $\pm$ 0.05\\
    {[Fe/H]} & -1.82 $\pm$ 0.10 & -1.81$^{+0.07}_{-0.06}$\\
    {[Cr/H]} & -1.71 $\pm$ 0.10 & -1.69 $\pm$ 0.07\\
    {[Si/H]} & -0.87 $\pm$ 0.10 & -0.86$^{+0.16}_{-0.10}$\\
    Dust-to-gas ratio, $\kappa$ & 0.10 $\pm$ 0.2 & 0.1\\ 
    Micro-turbulence & 2 $\pm$ 1 km/s & 19.3 km/s \\
    Cosmic ray ionization rate & -15.70 $\pm$ 0.10 log s$^{-1}$& -\\
    Magnetic field at illuminated face & 2.5 $\pm$ 0.6 $\mu$G & -\\
 \hline
\end{tabular}
\label{tab:DLA3a}
\end{table}
\end{comment}

\subsection{DLA towards PSS J1443+2724 ($z_{abs}=4.224$)}
The DLA at $z=4.224$, lying along the sightline to PSS J1443+2724 ($z_{em}=4.42$), was observed by \citet{Ledoux2006} using UVES at VLT, Chile. The wavelength coverage was 478-681 nm at a spectral resolution of 48,000 for the duration of the observation. The metallicity estimate was based on the value of [S/H] = -0.63, and the dust-to-gas ratio was found to be $\kappa$ = 0.15 respectively. This system was noted for its high metallicity ([Fe/H] = -1.12 $\pm$ 0.10) among the known $z > 4$ DLAs \citep{Ledoux2003}, and still remains the highest-redshift DLA known to host molecular hydrogen with H$_2$ arising from three different components \citep{Ledoux2006}. \citet{Ledoux2006} deduced the presence of cold gas with temperature $\sim$ 90-180 K and with a particle density, n$_{\rm H} \leq$ 25 cm$^{-3}$. We model the system as a single cloud by adding up the component-wise observed column densities. The H$_2$ J = 4 column density was accurately determined in one of the three components, with upper limits reported for the other two components. %We determine the sum of these observed column densities and use this as a limiting value for the J = 4 level in our models.

The gas density calculated by the {\sc cloudy} model for this cloud remains steady before rising sharply to 150 cm$^{-3}$ at $4 \times 10^{20}$ cm into the cloud, which happens to coincide with the region where the temperature drops from $\sim$ 10$^4$ K to 90 K. Given the parameter uncertainties, the gas density and temperature in the cloud centre spans 70--160 cm$^{-3}$ and 80--100 K, respectively. Thus, the gas in the central region of this DLA is relatively cold and dense. The variations in both trends as a function of depth into the cloud are shown on the left panel of Fig. \ref{fig:DLA_4.224}, alongside the pressure trends. The gas pressure within the cloud's interior is approximately 14,000 cm$^{-3}$ K, which falls within the range of 6600–15,000 cm$^{-3}$ K derived from our models spanning the uncertainty estimates of our input physical parameter values. This range is consistent with typical pressure values reported for H$_2$-DLAs \citep{Srianand2005}. As in previous cases, we find that thermal pressure surpasses turbulent pressure in the atomic phase. The bottom row of Fig. \ref{fig:DLA_4.224} shows the densities of ionic, atomic, and molecular hydrogen. Similar to the DLA at 3.255, the \ion{H}{i}-H$_2$ transition is not fully achieved, even in the cloud’s core. 

\ion{H}{i} photoionization serves as the dominant heating mechanism at shallower depths but is eventually overtaken by cosmic-ray heating beyond a distance of $10^{20}$ cm from the illuminated face of the cloud. Cooling due to collisions of H atoms peaks with a contribution of 80\% to the total cooling in the exteriors. Then its importance declines as the oxygen and silicon emission channels become more significant cooling agents in the deeper regions of the cloud.

\subsection{DLA towards Q J1236+0010 ($z_{abs}=3.033$)}

The DLA towards Q J1236+0010 was observed as part of a larger survey by \citet{Balashev2019} using the multi-wavelength, medium-resolution X-Shooter located on Kueyen at VLT, Chile. Unlike UVES, X-Shooter has a wavelength coverage of 300-2500 nm and a spectral resolution of 4000-17,000. \citet{Balashev2019} have obtained constraints on some physical parameters of the DLA based on the observed column densities of metals, H$_2$ and \ion{C}{i}. The metallicity of the absorber is determined to be -0.58$^{+0.04}_{-0.03}$ based on sulphur absorption lines, while the molecular fraction, f = -0.80$\pm$0.01. The study also inferred the total hydrogen density $\sim$ 10--100 cm$^{-3}$ corresponding to the molecular part of the DLA cloud, H$_2$ rotational excitation temperature H $_2$ $\sim$ 120 K, and the intensity of the incident UV radiation field between 2--11 G$_0$, where G$_0$ refers to the Habing field \citep{Habing1968, Tielens1985}.

\begin{comment}
\begin{table}
\centering
\caption{Input parameter values of the best-fitting model for the DLA towards Q J1236+0010 at $z_{abs} = 3.033$}
    \begin{tabular}{lcc}
    \hline
    Physical parameter & Model value & Observed value\\
    \hline
    Radiation field & KS19+CMB+ X-rays & -\\  
    UV Intensity of KS19 & 0.29 G$_0$ &-\\ 
    X-ray ionization parameter & -4.35 $\pm $ 0.05 & -\\
    Density at illuminated face, n$_H$ & 0.44 $\pm $ 0.10 log cm$^{-3}$ & 10-100 cm$^{-3}$ \\
    & & (in molecular region) \\
    Metallicity & -0.63 $\pm $ 0.10 & -0.58 $^{+0.04}_{-0.03}$\\
    {[C/H]} & -2.21 $\pm $ 0.15 & -\\
    {[O/H]} & -1.03 $\pm $ 0.05 & -\\
    {[Si/H]} &  -0.98 $\pm $ 0.05 & -\\
    %{[S/H]} & -0.63 $\pm $ 0.02 & -0.58 $\pm$ 0.04\\
    {[Fe/H]} & -2.13 $\pm $ 0.05 & -1.09 $\pm$ 0.04\\
    Dust-to-gas ratio, $\kappa$ & 0.30 $\pm $ 0.04 & 0.24 \\
    Micro-turbulence & 2.3 $\pm$ 0.2 km/s & 2.3 $\pm$ 0.03 km/s \\
    Cosmic ray ionization rate & -15.70 $\pm$ 0.10 log s$^{-1}$ & - \\
    \hline
    \end{tabular}
    \label{tab:DLA4a}
\end{table}
\end{comment}

We model this system using an incident intensity of 2.54 log G$_0$, incident total hydrogen density of 2.97 log cm$^{-3}$ and cosmic ray ionisation rate of -14.01 log s$^{-1}$. In Fig. \ref{fig:3.033_physical}, we see the variation in the physical state of the absorber with depth from the illuminated face. The total hydrogen density increases while the temperature decreases. This shift becomes prominent at the photodissociation front, which in our model happens to be at about $1.6 \times 10^{16}$ cm. In the interior of the cloud, the total hydrogen density and the gas temperature reach values of $(3.7 - 4.3) \times 10^4$ cm$^{-3}$ and 85--87 K, respectively. The gas pressure drops from the surface layers of the DLA model to the interiors and reaches a value within $(2.8-3.2) \times 10^6$ cm$^{-3}$ K in the molecular regions. As seen in the bottom left panel of Fig. \ref{fig:3.033_physical}, the molecular form of hydrogen becomes the most dominant form of the species at depths beyond this transition point. The level population can be noted in the adjacent panel -- more H$_2$ exists at these depths in the lower levels and the number of molecules in higher rotational states falls away due to the weaker attenuated radiation field coupled with the cooler and denser environment. 

Heating due to \ion{H}{i} photoionization is significant only at shallower depths in the cloud, whereas grain photoelectric heating and cosmic rays are significant at depths where the gas has started harbouring a significant fraction of molecules. Cooling in the inner regions is dominated by the [\ion{C}{ii}] 158 $\mu$m, [\ion{O}{i}] 63 $\mu$m line emission, while the [\ion{Si}{ii}] 35 $\mu$m emission and radiation following collisional excitation within the ground electronic state of H$_2$ also play important roles.

\section{DISCUSSION}
\label{sec:disc}

%Through the models described in the previous section, we successfully reproduce $\sim$ 87 per cent of all the available observed column densities for the six DLAs to within a factor of two, or to within observational error, depending on which is larger.

We have successfully modelled six of the highest-redshift H$_2$-DLAs through a combination of {\sc cloudy} model grids and Bayesian analysis. We discuss the results here. Four of these systems have been previously modelled by others with some differences in methodology. Hence, we also provide a comparison between our results and those available in the literature for these four DLAs.

\subsection{Physical environment of the modelled DLAs}
\label{ssec:phy-envir}

All our DLA models are impinged by the metagalactic background radiation \citep{2019} along with the cosmic microwave background, both of which have a redshift dependence. In addition to this, we constrain the contribution of the local interstellar radiation field. While the DLA at z = 4.224 is subject to a low-intensity ISRF comparable to the intensity of the Habing field \citep{Habing1968}, the other DLAs are irradiated by much stronger fields. The UV radiation incident on the DLAs at z = 3.091 and 3.287 correspond to $\sim$ 35 and 102 G$_0$ respectively, while for the DLAs at 3.033, 3.093, and 3.255, the field strength is higher, at 346, 602, and 467 G$_0$. This points to significant \textit{in situ} star formation in these DLA environments. Though DLAs have typically been associated with low star-formation activity \citet{Rahmani2010}, our estimates of I$_{UV}$ are mostly consistent with results from other H$_2$-DLA studies such as those by \citet{Klimenko2020} and \citet{Kosenko2021}.

The mean cosmic ray ionization rate of neutral hydrogen in the Galaxy was constrained by \citet{Indriolo2007} to be 2 $\times$ 10$^{-16}$ s$^{-1}$. Over the years, several observational and modelling-based estimates have been obtained for Galactic sightlines \citep{McCall2003, Shaw2008, Indriolo2012, Obolentseva2024}. It still remains an active area of research. Even less is known about the cosmic ray ionization rate at high redshift. \citet{Indriolo2018} used OH$^+$ and H$_2$O$^+$ to estimate the cosmic ray ionization rate in the gaseous halos associated with lensed galaxies at z $\sim$ 2.3. They found the value to lie in the range 10$^{-16}$ to 10$^{-14}$ s$^{-1}$. Our results are consistent with this estimate. Only the rate values constrained for the DLAs at z = 3.033 lie close to the upper limit of this range. 

The typical density and temperature conditions in the WNM are $\sim$ 0.6 cm$^{-3}$ and 5000 K, while in the CNM, they take values $\sim$ 30 cm$^{-3}$ and 100 K \citep{Draine2011}. In diffuse molecular gas, the density may be higher, $\sim$ 100 cm$^{-3}$, while the temperature can reach up to $\sim$ 50 K. In denser regions, the temperature is lower still while density can increase beyond 1000 cm$^{-3}$. In all our DLA models, the gas exhibits multiple phases. The temperatures in the interior regions are typical of the CNM and diffuse molecular gas. As our models have constant pressure across the cloud, the density varies at different depths, rising in the inner parts of the cloud. These values are mostly higher than conventional expectations of cool molecular gas. However, such values ranging from 300 to 4500 cm$^{-3}$ are still mostly consistent with interstellar conditions in the Galaxy \citep{Federman1983, Gerin2015}.

The median gas pressure in the Galactic ISM is $\sim$ 3800 cm$^{-3}$ K. \citet{Jenkins2011} observationally constrained values ranging from 100 to 10$^{5}$ cm$^{-3}$ K. In high-redshift H$_2$-DLAs, the thermal pressure was estimated to lie between 824 and 30,000 cm$^{-3}$ K \citep{Srianand2005}. While most models are consistent with these estimates in the cooler parts of the cloud, the DLAs at z = 3.033 and 3.255 harbour gas pressure $\sim$ 10$^5$ K. Higher pressures are typically expected in molecular cloud environments and regions associated with star-formation \citep{Anathapindika17}. \citet{Jenkins2001} also discuss the existence of higher-pressure regions in the ISM, but in small localized regions. A large range of pressure ($10-10^6$ K cm$^{-3}$) is allowed within the two-phase model of \citet{Wolfe2003b} for DLAs. \citet{Neeleman2015} follow along the same lines to suggest that their eight DLAs trace a highly turbulent ISM in young, star-forming galaxies, with pressures exceeding 20,000 K cm$^{-3}$.

H$_2$-DLAs are generally considered to be small, typically $\leq$ 15 pc, based on radio observations of twenty-eight DLAs that lacked H$_2$ absorption \citep{Srianand2012}. In comparison, diffuse neutral clouds just outside the Galactic plane have diameters ranging from 10 to 100 pc \citep{Pidopryhora2015}. From the model results, we infer that the DLAs at z = 3.033, 3.091, 3.255, 3.287,  4.224, and 3.093 have longitudinal extents of 0.02, 0.29, 0.5, 31, 96, 221 and 264 pc respectively. In terms of $A_V$, this corresponds to 0.05, 0.003, 0.2, 0.5, 0.06, and 1.21. Thus, two of our modelled DLAs are extremely small. But similar sizes for DLAs have been reported by \citet{Jorgenson2009} and \citet{2020Balashev}. The largest cloud among all our DLA models is the absorber at z = 3.093. However, this is still consistent with the extent of translucent interstellar clouds which corresponds to $A_V = 1-2$ \citep{Snow2006}. 

The density and pressure in the DLA towards QJ1236+0010 at $z_{abs}$=3.033 are higher than that of diffuse gas that is considered typical of the circumgalactic medium, though the temperature in the innermost part of the absorber does drop below 100 K. The inferred pressure value of $\log(P/k) = 6.6$ cm$^{-3}$ K in this DLA is consistent with the ISM pressure observed in \ion{H}{ii} regions in the Large and Small Magellanic Clouds, which range from $\log(P/k) = 6$ to 7.5 cm$^{-3}$ K \citep{2023Jin}. Additionally, it falls within the broader ISM pressure range of $4 < \log(P/k) < 9$ cm$^{-3}$ K predicted by the theoretical models of \citet{2019Kewley} who studied local and high-redshift galaxies. While the $n_H$ predicted in the interiors of our best-fitting model for this DLA is higher than the values of few hundred cm$^{-3}$ reported for H$_2$-bearing DLAs in \citet{Srianand2005}, such high densities are not entirely unprecedented in DLAs. \citet{Noterdaeme2019} demonstrated that the distance of a cloud from the background quasar influences the density required for the \ion{H}{i}-H$_2$ transition. In case of proximate DLAs, that is DLAs which are located close to the host quasar, significantly higher densities ($n_H \sim 10^4$ cm$^{-3}$) are required to form H$_2$. As a consequence, they are also expected to have small physical dimensions of less than 0.1 pc. This is clearly the case with the DLA at $z = 3.033$, which has been identified in literature as a proximate DLA. Similarly, the DLA towards Q J2100-0600 at $z_{abs} =3.091$, also falls into this category and is the second smallest DLA modelled in our sample, with the hydrogen densities being possibly as high as 2000 cm$^{-3}$.

Many H$_2$-DLAs have been observationally studied at redshifts in the range 2--3 \citep{Ledoux2003, Noterdaeme2008}. As a consequence, most detailed models that give insight into the physical conditions of DLAs also correspond to this redshift interval \citep{Shaw2016, Noterdaeme2017, Balashev2017, Rawlins2018, Klimenko2020}. A key aim of the present study was to understand the physical environment of DLAs that lie beyond this redshift. As discussed before, our models constrain high estimates of ISRF and cosmic ray ionization rates for some of the modelled sightlines. However, given the limited number of sightlines, it is not possible to draw any robust conclusions about any evolutionary trends in the properties of DLAs. More observations and models of DLAs at similar redshifts will enable more detailed probing of these distant environments, and help understand any evolutionary patterns in the physical conditions extant in DLAs across all redshifts.

\subsection{Comparison with other models}
\label{ssec:model-comp}

Four of the DLAs we modelled in this work have been previously studied. One of them, at z = 3.093, was analyzed by \citet{Kosenko2021} to investigate the presence of HD, and another one by \citet{Balashev2019}, while the other two, at z = 3.091 and z = 3.28, were modelled by \citet{Klimenko2020} using the {\sc meudon-pdr} code \citep{2006LePetit}.

A significant feature of our modelling approach is that we consider a constant pressure system that imitates the pressure equilibrium expected in media where multiple gas phases (that is, both WNM and CNM) co-exist. In comparison, most other models in the literature consider constant density across the absorbing cloud. Thus, our models span the entire extent of the DLA and we constrain them using all observed atomic and molecular species distributed across the warm and cool phases along the sightline.

\citet{Klimenko2020} considered a slab model exposed to an isotropic ISRF as defined by \citet{1983MMP} from both sides. Cosmic ray ionization rate and gas turbulent velocity were held constant at $10^{-16}$ s$^{-1}$ and 2 km/s respectively, while the cloud size was specified using the visual extinction condition, A$_{V}^{max} = 0.5$. In addition, their model incorporates polycyclic aromatic hydrocarbons and a slightly wider range of grain sizes (0.001-0.3 $\mu$m) than adopted in our models. We execute in {\sc cloudy}, models using the parameters specified by \citet{Klimenko2020}, and reproduce low-H$_2$ and \ion{C}{i} column densities consistent with observations within about 0.3 dex. The modelling approach by \citet{Klimenko2020} uses the constant-density approach and relies mostly on reproducing the lowest rotational levels of H$_2$ (J $\leq$ 2) and the fine structure levels of \ion{C}{i}, that is, they model only the cool molecular region of the gas.

Here, we compare the results of our modelling approach with that of \citet{Klimenko2020}. The absorber at z = 3.28 was modelled by \citet{Klimenko2020} with an intensity of log I$_{UV} = -0.32^{+0.20}_{-0.16} $ (in Mathis units), density value of log n$_H = 1.63^{+0.08}_{-0.12}$ cm$^{-3}$ and average metallicity of [X/H] = –1.10 $\pm$ 0.10. In comparison, we deduce log I$_{UV}$ = 2.01$\pm 0.03$, that is, 1.97 (in log Mathis units), log n$_H$ = 1.03 $\pm 0.04$ cm$^{-3}$ at the illuminated face of the constant-pressure cloud, and log $\zeta$ = -16.99 $\pm 0.01$ s$^{-1}$. We use [X/H] = -1.10 based on observational estimates. Further, the density increases to log n$_H$ = 2.90$_{-0.61}^{+0.24}$ cm$^{-3}$ in the interior of the cloud. As our model maintains constant pressure across the cloud, the density in the interior traces denser gas, along with a much enhanced intensity at the incident part of the cloud. 

For the DLA at z = 3.091, \citet{Klimenko2020} constrain values of log I$_{UV}$, log n$_H $ and [X/H] to be  $-0.50^{+0.22}_{-0.34}$ (in log Mathis units), $1.40^{+0.28}_{-0.35}$ (in log cm$^{-3}$), and $-0.73 \pm 0.15$, respectively. In our model,  we obtain log I$_{UV}$ = $1.55^{+0.02}_{-0.03}$ (1.46 in log Mathis units), density at the incident face of the cloud to be $1.25^{+0.27}_{-0.11}$ (in log cm$^{-3}$) and log $\zeta= -14.50^{+0.42}_{-0.03}$. In the cloud interiors, the density increases to $2.88_{-0.10}^{+0.42}$ (in log cm$^{-3}$). We set [X/H] = -0.73 in our model as per observational deduction.

The H$_2$-bearing part of the absorber at z = 3.033 was modelled by \citet{Balashev2019} with an intensity of log I$_{UV} = 0.40^{+0.40}_{-0.40} $ (in Draine units), density value of log n$_H = 1.5^{+0.60}_{-0.60}$ cm$^{-3}$ and average metallicity of [X/H] = –0.58$^{+0.04}_{-0.03}$. In comparison, we deduce log I$_{UV}$ = $2.54 \pm 0.02$, that is, 2.39 (in log Draine units), log n$_H$ = 2.97$_{-0.04}^{+0.02}$ cm$^{-3}$ at the illuminated face of the constant-pressure cloud, and log $\zeta$ = -14.01$_{-0.02}^{+0.01}$ s$^{-1}$. We use [X/H] = -0.58 based on observational estimates.
%PDR Meudon uses a default of 10^-17,
%Given the notable differences in the methodology adopted by \citet{Klimenko2020} compared to ours, some differences in parameter values are expected. Despite these differences, our results generally align well, with most values falling within comparable ranges. The main deviation is evident in the incident UV intensity, where our values are higher, primarily due to the inclusion of the warmer phase of the clouds in our modeling.  

\citet{Kosenko2021}'s investigation employed a semi-analytic approach to describe the dependence of the HD/H$_2$ ratio on physical parameters, based on the balance between HD formation and destruction in a plane-parallel, steady-state cloud. The DLA at z = 3.093 towards J1311+2225 exhibits H$_2$ absorption in 4 components. The physical parameter values constrained by them for component 2, that is the most molecular component, are as follows: log I$_{UV}$ = 1.1$^{+0.1}_{-0.1}$ (with respect to the Draine field), log n$_H$ = 1.7$^{+0.2}_{-0.2}$ cm$^{-3}$, and log $\zeta$ = -16.2$^{+0.1}_{-0.1}$ s$^{-1}$. Unlike \citet{Klimenko2020}, the cosmic ray ionization rate is not held fixed and is one of the parameters constrained here.
In comparison, our models and analysis predict log I$_{UV}$ = 2.78$^{+0.12}_{-0.26}$, that is, 2.63 with respect to the Draine field, log n$_H$ = 0.62$^{+0.05}_{-0.02}$ cm$^{-3}$, and log $\zeta$ =  -14.54$ \pm 0.03$ s$^{-1}$. Here too, our model predicts a larger cosmic ray ionization rate, while density varies between 3-30 cm$^{-3}$ across different regions of the absorber.

The main difference between our model results as compared to those previously reported in literature lies in the predicted interstellar radiation field strength and cosmic ray ionization rate. Isobaric models such as the ones we have set up here, have more sound physical motivation than models assuming constant density. So we emphasise the necessity of adopting an appropriate modelling framework, as that itself can be an influential factor in accurately constraining the physical properties of an absorber.

\section{SUMMARY}
\label{sec:conc}

Probing the physical environment of DLAs at various redshifts is vital to our understanding of how galaxies and their environments interact and evolve over time. The current work focuses on deriving the physical properties of the most distant H$_2$-DLAs (redshifts in the range 3--4.5). The six systems we study lie at redshifts 3.033, 3.091, 3.093, 3.255, 3.287 and 4.224. We used the software {\sc cloudy} to model the environment of these DLAs by setting up grids of models spanning the typical range of conditions seen in circumgalactic gas, and then subjecting these models to Bayesian analysis techniques. 
%by matching the predicted column densities of various species to within a factor of 2 of the observed values. 
The key takeaways from our study are as follows:
\begin{enumerate}
    \item We implement an approach of maintaining constant pressure throughout our model cloud, as compared to most modelling studies which assume constant density. This allows us to model the entire multiphase DLA and use all observed atomic and molecular species as constraints to build a more efficient model.
    \item The six DLAs are irradiated by the metagalactic background and cosmic microwave background at the appropriate redshift. The models also included an additional component of UV radiation from the interstellar radiation field. The intensity of this radiation field ranges from 10 to $10^3$ G$_0$, where G$_0$ is the mean Galactic interstellar radiation field. This suggests \textit{in situ} star-formation, and so, some of these DLA environments are likely associated with significant star-formation activity. 
    %\item The DLAs at z = 3.033 and 3.091 are subject to an enhanced X-ray radiation field extending from 1--100 keV. Thus, these are inferred to be X-ray dominated regions.
    %\item In order to replicate the relatively large column densities in the higher rotational levels of H$_2$ in the absorbers at redshifts $z=3.255$ and 4.224, it is necessary to include magnetic fields of strength 2--9 $\mu$G, respectively. The magnetic field helps in significantly increasing the population of the H$_2$ rotational levels, making them comparable to observed column densities. Magnetic field pressure is dominant in the interior of these clouds.
    %\item Alternately, we find that the observed H$_2$ level population for the DLAs at z = 3.255 and 4.224 could potentially be modelled by accounting for additional heating within the absorber. As such heating could be the result of mechanical effects, we conclude that these DLA environments can only be understood fully by adopting an appropriate hydrodynamic/magnetohydrodynamic approach.
    \item The modelled DLAs constitute both diffuse, warm gas as also denser cold gas. The density, temperature and pressure in the interior of the various clouds lie in the range 50--40,000 cm$^{-3}$, 35--200 K  and 10$^{3.5}$ to 10$^{6.4}$ cm$^{-3}$ K respectively. While the density and temperature values are consistent with the expectations for the WNM and CNM, the thermal pressure is higher than typically constrained for high-redshift H$_2$-DLAs.
    \item The cosmic ray ionization rate was found to span a wide range of values, from 10$^{-17}$ to 10$^{-14}$ s$^{-1}$ for the six DLAs. The lower end of these values is consistent with conditions in diffuse atomic and molecular clouds in the Galaxy. Though rates significantly higher than 10$^{-15}$ s$^{-1}$ are not very common, they are consistent with recent estimates for high-redshift environments.
    %\item The DLA models are less than 15 pc across. We find that the longitudinal extent of the other DLA clouds spans a range of $\sim$ 50 to 300 pc. This is larger than the sizes predicted by models of DLAs which lie within the redshift interval 2--3. But they are consistent with observational estimates of DLA extent made based on impact parameters with respect to the associated galaxy. \textcolor{cyan}{update}
    \item Heating in these DLAs is dominated by \ion{H}{i} and \ion{He}{ii} photoionization, cosmic rays, and the photoelectric heating of grains. The specific depth at which each heating mechanism dominates—whether at shallow, intermediate, or deeper regions of the cloud—varies depending on the individual system. Cooling in the interior regions occurs through a combination of [\ion{C}{ii}] 158 $\mu$m, [\ion{O}{i}] 63 $\mu$m and [\ion{Si}{ii}] 35 $\mu$m emission.
    %\item CII and IR luminosity, if explored
    \end{enumerate}

\section*{Acknowledgements}
The authors are thankful to the anonymous referee for insightful comments that have improved the quality of the manuscript. They also wish to thank Ksenia Telikova for sharing some of the unpublished observed column densities for the DLA at z = 3.255. AAS would like to thank Christophe Morisset for his inputs on uncertainty calculations using CLOUDY, and Romain Meriot and Harshank Milind Nimonkar for their help with Bayesian statistics. KR is grateful to Bhaswati Mookerjea for helpful discussions about the physical conditions in interstellar gas. Software used for this work: {\sc emcee} \citep{2013Foreman-Mackey}, Python \citep{vanRossum09}, Numpy \citep{Harris20}, Matplotlib \citep{Hunter07}, Scipy \citep{Virtanen20}, Astropy \citep{Price-Whelan22}, and Pandas \citep{Pandas22}.

%%%%%%%%%%%%%%%%%%%%%%%%%%%%%%%%%%%%%%%%%%%%%%%%%%
\section*{Data Availability}
No new observations are presented in this paper. Further details about the {\sc cloudy} models or their results can be shared upon reasonable request to the authors.

%%%%%%%%%%%%%%%%%%%% REFERENCES %%%%%%%%%%%%%%%%%%

% The best way to enter references is to use BibTeX:

\bibliographystyle{mnras}
\bibliography{example} % if your bibtex file is called example.bib

\appendix
\section*{APPENDIX}
\renewcommand{\thefigure}{A\arabic{figure}} % Prefix figures with "A"
\setcounter{figure}{0} % Reset figure counter for appendix

In this appendix, we present the corner plots used to derive maximum-probability value along with the uncertainty estimates for key input parameters such as hydrogen density ($n_H$), ionizing ultraviolet radiation ($I_{UV}$), and cosmic ray ionization rate ($\zeta$). Additionally, we include the physical conditions plots, which illustrate the variation of parameters such as density, temperature, pressure constituents, and abundances as a function of depth into the cloud.

\begin{figure}
    \centering
    \includegraphics[width=\linewidth]{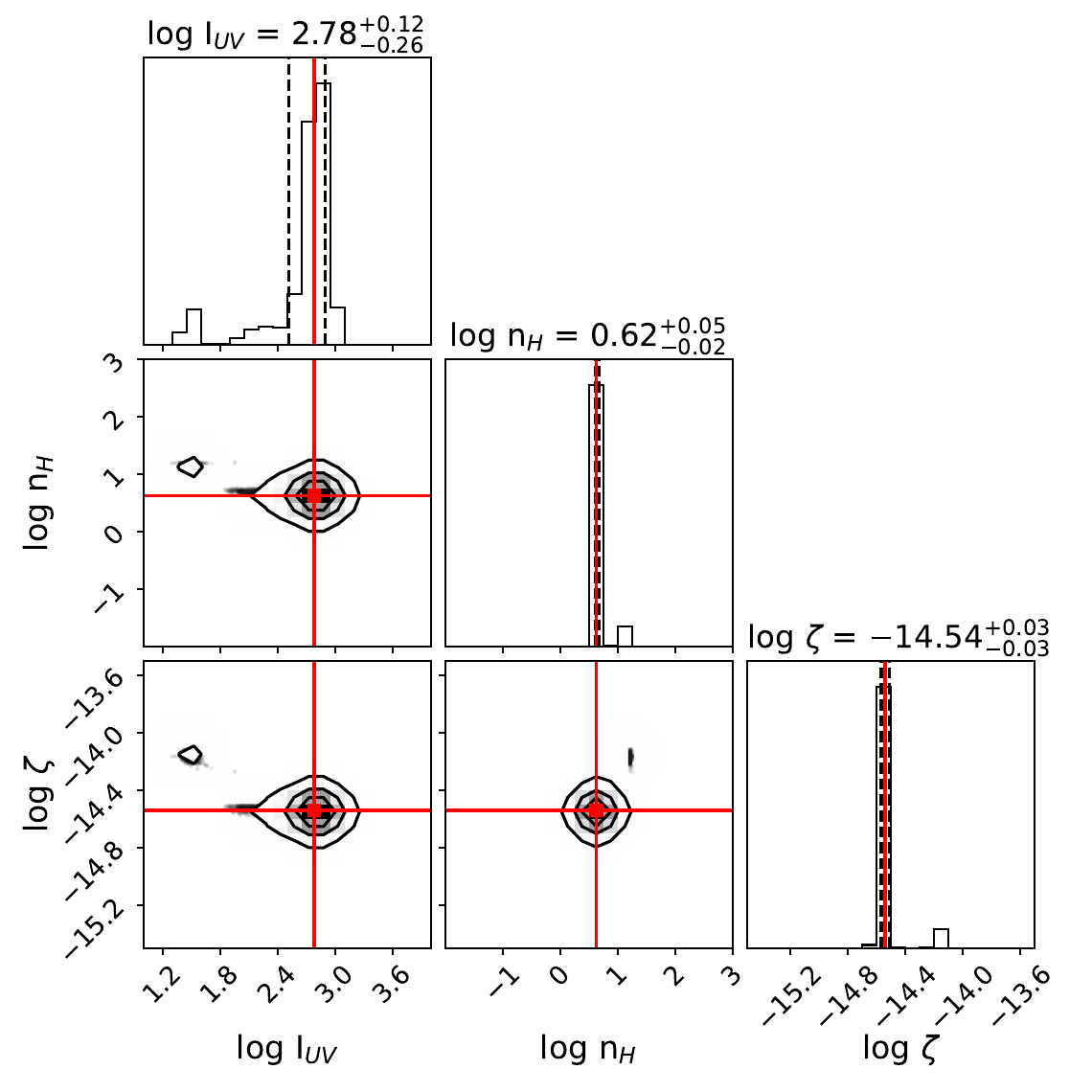}
    \caption{Same as Fig. \ref{fig:corner_3.287} for DLA towards J1311+2225 at $z_{abs}=3.093$. }
    \label{fig:corner_3.093}
\end{figure}

\begin{figure}
    \centering
    \includegraphics[width=\linewidth]{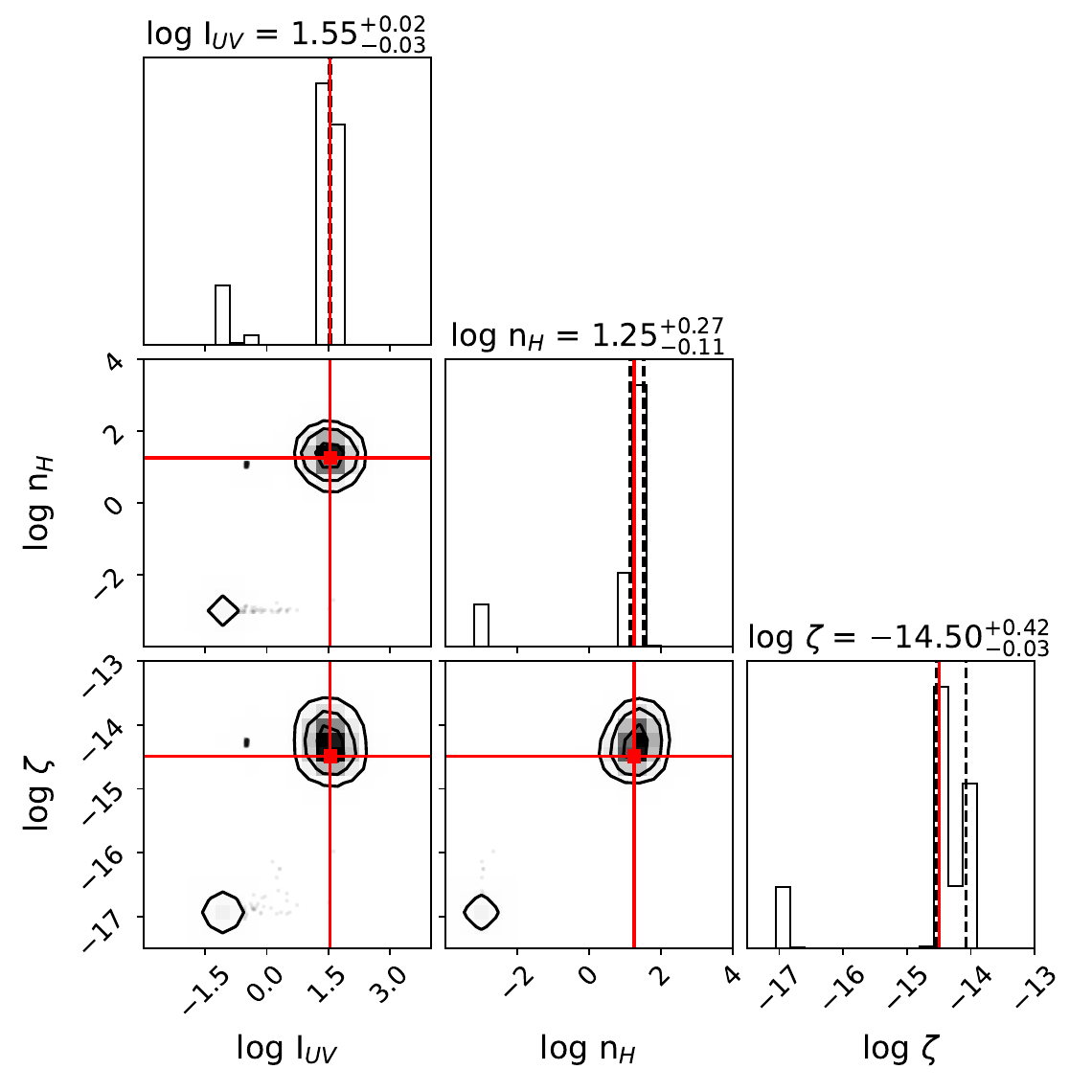}
    \caption{Same as Fig. \ref{fig:corner_3.287} for DLA Q J2100-0600 at $z_{abs}=3.091$.} 
    \label{fig:corner_3.091}
\end{figure}

\begin{figure}
    \centering
    \includegraphics[width=\linewidth]{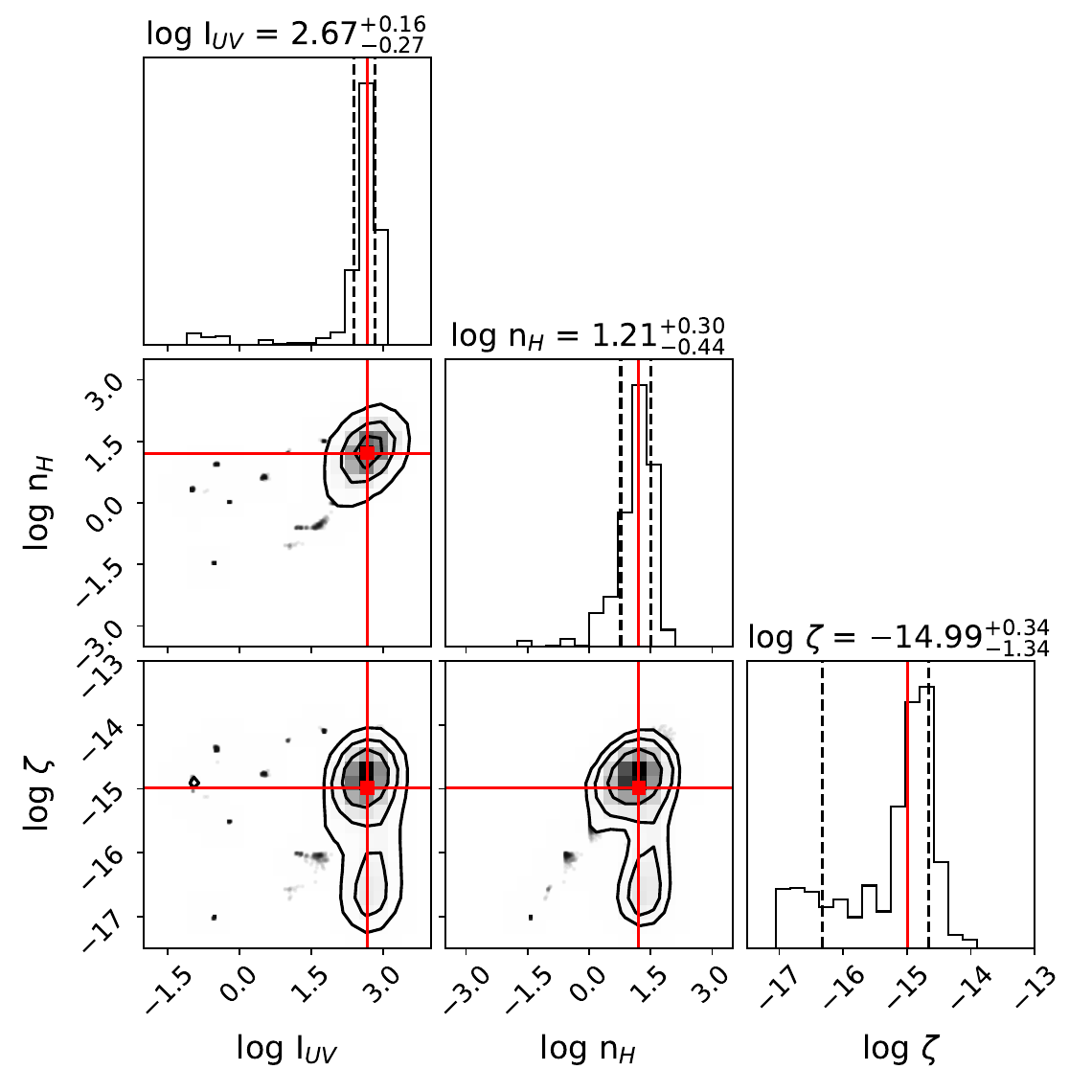}
    \caption{Same as Fig. \ref{fig:corner_3.287} for DLA J2205+1021 at $z_{abs}=3.255$.}
    \label{fig:corner_3.255}
\end{figure}

\begin{figure}
    \centering
    \includegraphics[width=\linewidth]{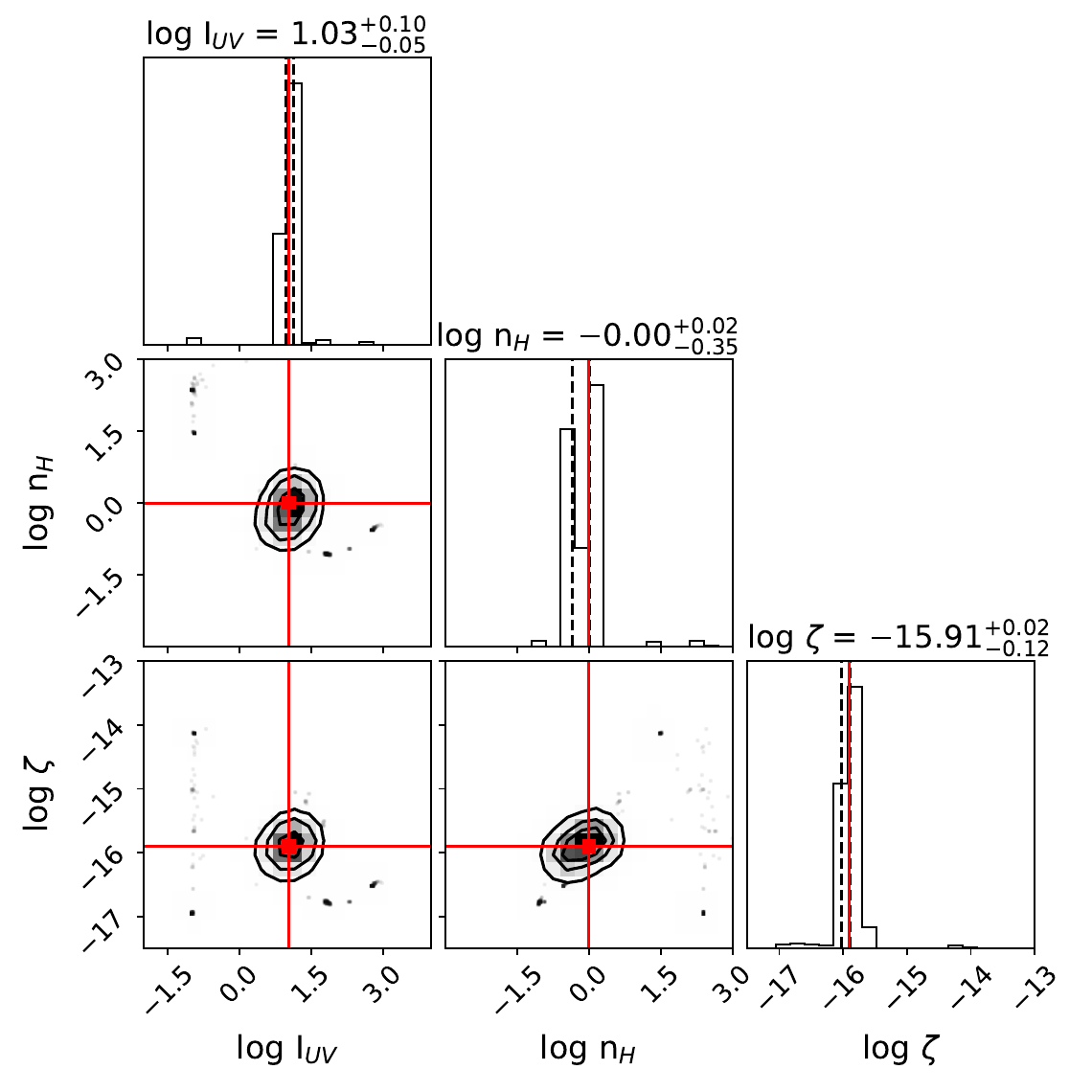}
    \caption{Same as Fig \ref{fig:corner_3.287} for the DLA towards PSS J1443+2724 at $z_{abs}=4.224$.} 
    \label{fig:corner_4.224}
\end{figure}

\begin{figure}
    \centering
    \includegraphics[width=\linewidth]{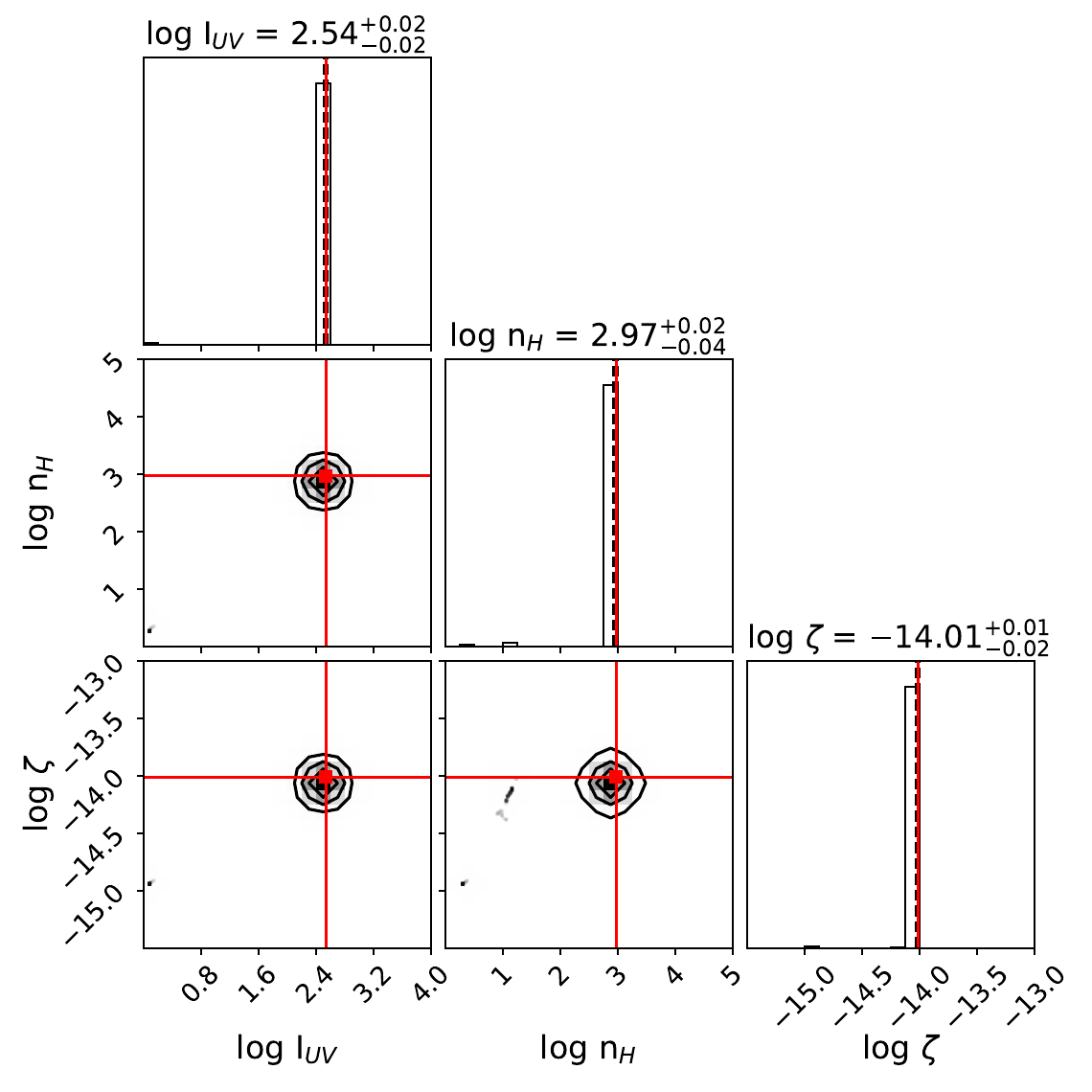}
    \caption{Same as Fig. \ref{fig:corner_3.287} for DLA towards Q J1236+0010 at $z_{abs}=3.033$.}
    \label{fig:corner_3.033}
\end{figure}

\begin{figure*}
    \centering
    \includegraphics[width=\linewidth]{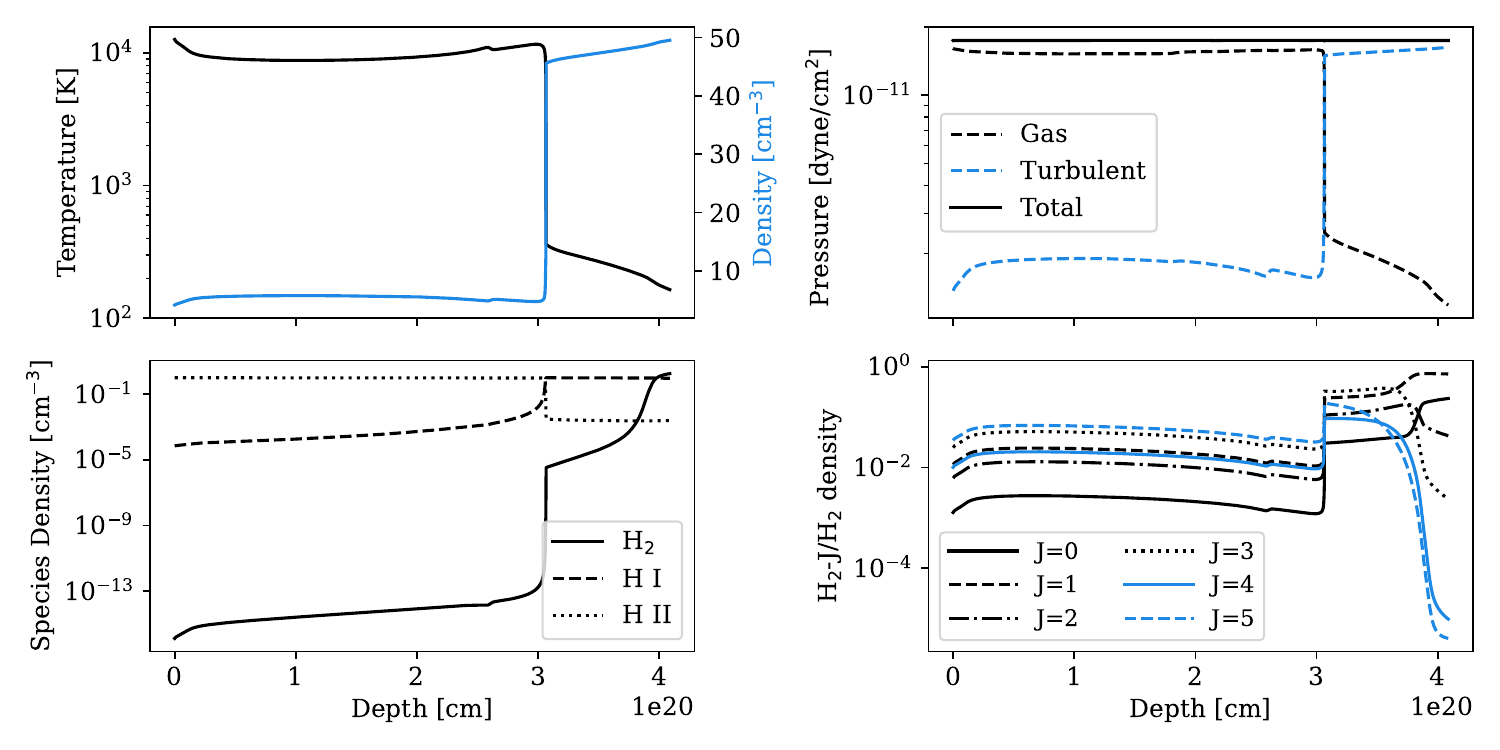}
    \caption{Same as Fig. \ref{fig:DLA1} for DLA towards J1311+2225 at $z_{abs}=3.093$. The input values of intensity, total hydrogen density, and cosmic ray ionisation rate for this model are 2.78 ($\log$ G$_0$), 0.62 ($\log \text{cm}^{-3}$), and -14.54 ($\log \text{s}^{-1}$).}
    \label{fig:3.093_physical}
\end{figure*}

\begin{figure*}
	\includegraphics[width=\textwidth, keepaspectratio]{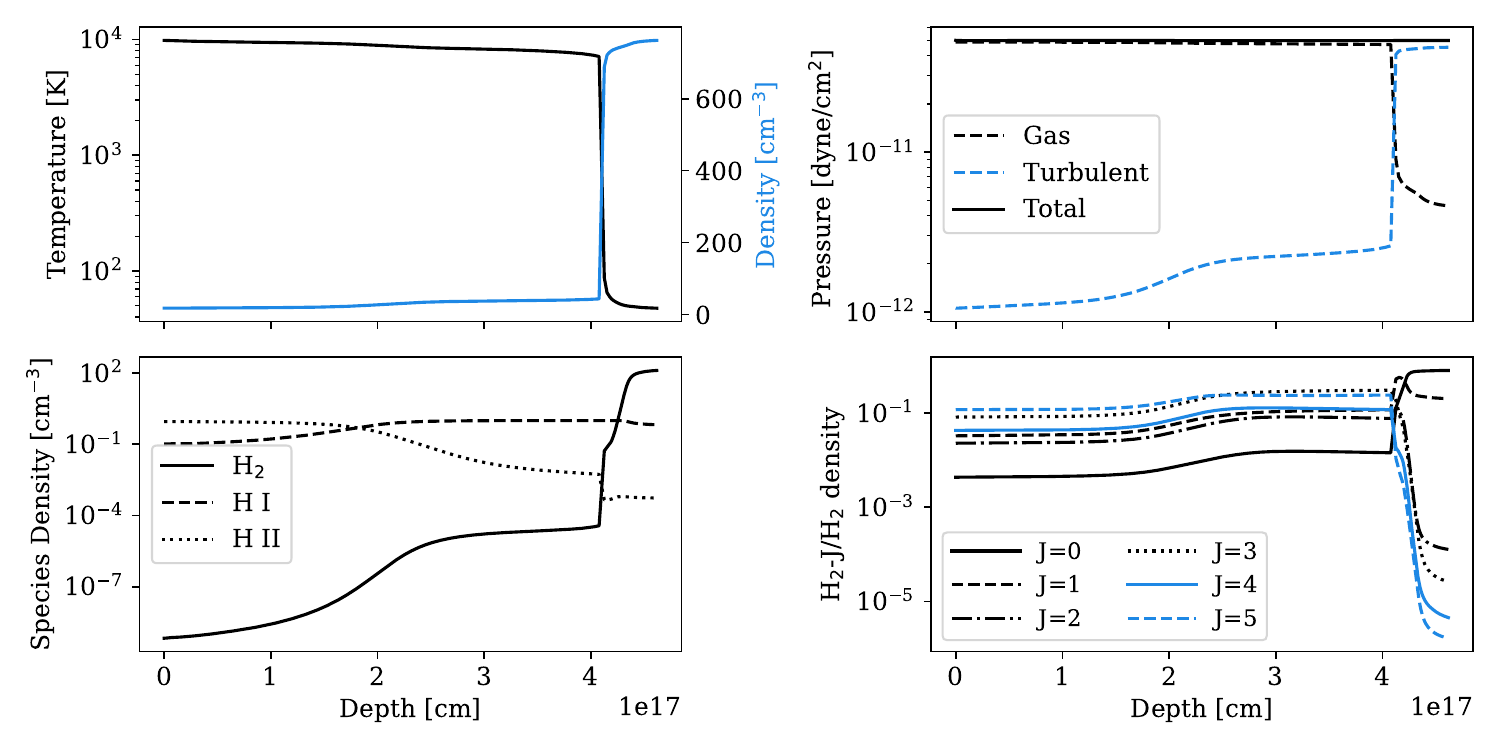}
    \caption{Same as \ref{fig:DLA1} for the DLA Q J2100-0600 at $z_{abs} = 3.091$. The input parameters for this model are $\log I_{UV} = 1.55$ G$_0$, $\log n_H = 1.25$ cm$^{-3}$, and $\log \zeta =-14.50$ s$^{-1}$.}
    \label{fig:DLA_3.091}
\end{figure*}

\begin{figure*}
	% To include a figure from a file named example.*
	% Allowable file formats are eps or ps if compiling using latex
	% or pdf, png, jpg if compiling using pdflatex
	\includegraphics[width=\textwidth, keepaspectratio]{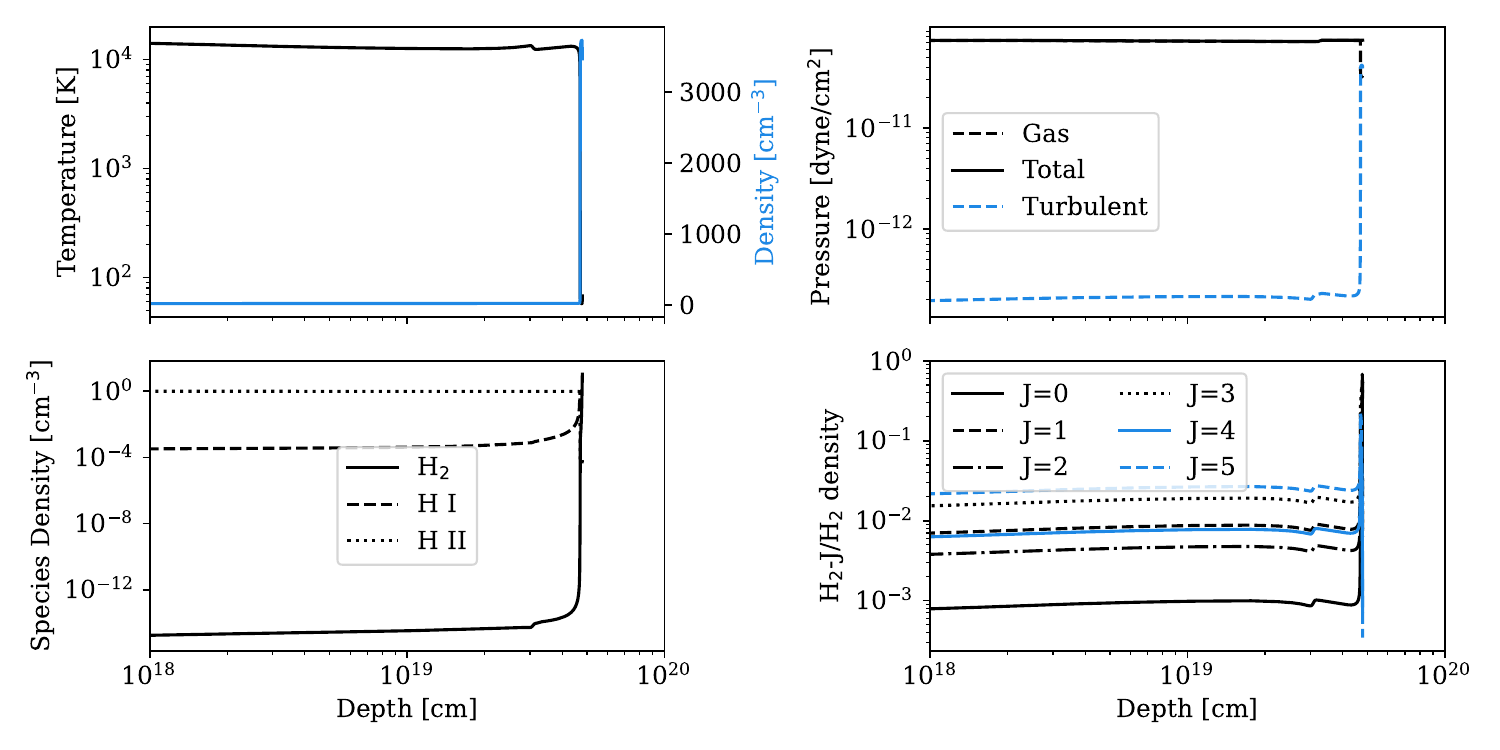}
    \caption{Same as Fig. \ref{fig:DLA1} for DLA J2205+1021 at $z_{abs}=3.255$. The input values of intensity, total hydrogen density, and cosmic ray ionisation rate for this model are 2.67 ($\log$ G$_0$), 1.21 ($\log \text{cm}^{-3}$), and -14.99 ($\log \text{s}^{-1}$).}
    \label{fig:3.255_physical}
\end{figure*}

\begin{figure*}
	\includegraphics[width=\textwidth, keepaspectratio]{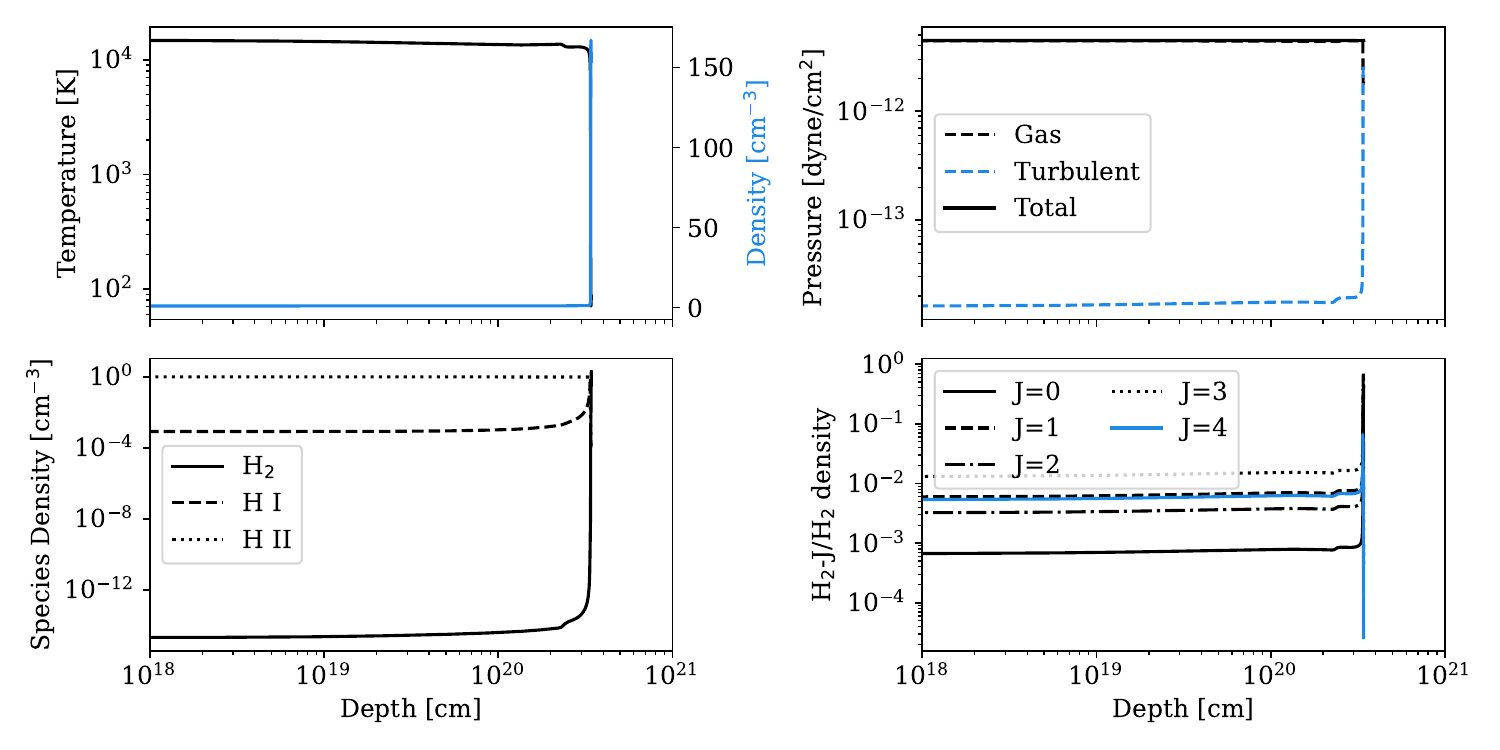}
    \caption{Same as Fig. \ref{fig:DLA1} for the DLA cloud at $z_{abs} = 4.224$ towards PSS J1443+2724. The input values of intensity, total hydrogen density, and cosmic ray ionisation rate for this model are 1.03 ($\log$ G$_0$), 0.00 ($\log \text{cm}^{-3}$), and -15.91 ($\log \text{s}^{-1}$).} 
    \label{fig:DLA_4.224}
\end{figure*}

\begin{figure*}
	\includegraphics[width=\textwidth, keepaspectratio]{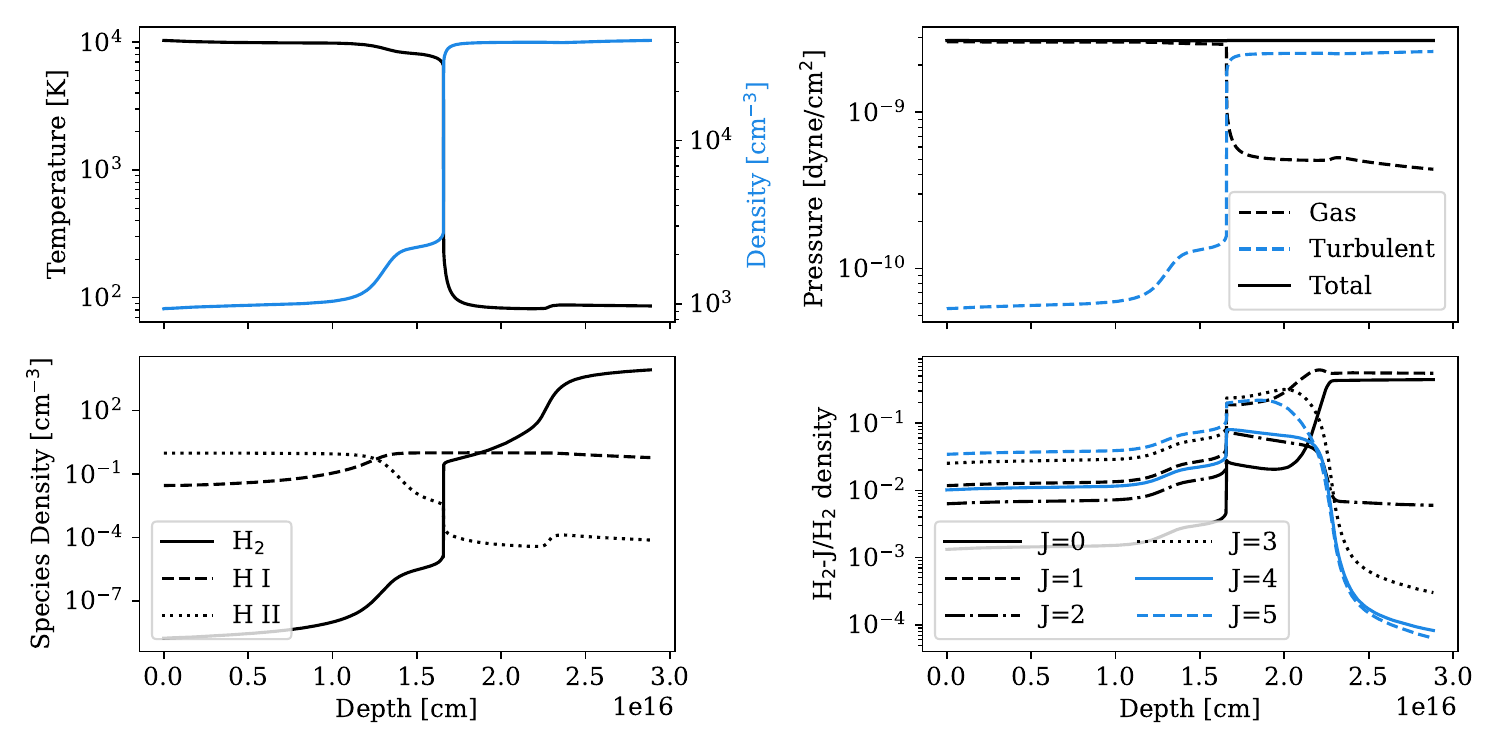}
    \caption{Same as Fig. \ref{fig:DLA1} for DLA towards J1236+0010 at $z_{abs} = 3.033$. The input values of intensity, total hydrogen density, and cosmic ray ionisation rate for this model are 2.54 ($\log$ G$_0$), 2.97 ($\log \text{cm}^{-3}$), and -14.01 ($\log \text{s}^{-1}$).}
    \label{fig:3.033_physical}
\end{figure*}

\bsp	% typesetting comment
\label{lastpage}
\end{document}